\documentclass[twocolumn]{aastex62}
\pdfoutput=1 
\usepackage{amsmath,amstext}
\usepackage[T1]{fontenc}
\usepackage[figure,figure*]{hypcap}

\usepackage{soul}
\graphicspath{{./}{figures/}}
\usepackage{CJKutf8}
\newcommand{\cjk}[1]{\begin{CJK}{UTF8}{gbsn}#1\unskip\end{CJK}\unskip}

\newcommand{\vsepHead}{\smash{\rule[-0.2ex]{0.4pt}{1.6ex}}}

\newcommand{\vsepDoubleTopGap}{\smash{\rule[-3ex]{0.4pt}{4.6ex}}}
\newcommand{\vsepDoubleBottomGap}{\smash{\rule[0.2ex]{0.4pt}{4.6ex}}}
\newcommand{\vsepDouble}{\smash{\rule[-0.9ex]{0.4pt}{4.8ex}}}    

\shorttitle{The Cosmic Shoreline, Sandbar, and an Airless Valley}
\shortauthors{Nguyen et al.}

\begin{document}

\title{An Evolving Cosmic Shoreline and Sandbar Bounding the Rocky Airless Valley}

\author[0009-0000-5587-7297]{Barron K. Nguyen}
\affiliation{Department of Earth and Planetary Sciences, Stanford University, Stanford, CA 94305-2115, USA}

\author[0000-0003-2915-5025]{Laura K. Schaefer}
\affiliation{Department of Earth and Planetary Sciences, Stanford University, Stanford, CA 94305-2115, USA}

\author[0000-0002-1592-7832]{Xuan Ji \texorpdfstring{(\cjk{纪璇})}{}}
\affiliation{Department of the Geophysical Sciences, The University of Chicago, Chicago, IL 60637, USA}

\author[0000-0002-9807-5435]{Christopher A. Theissen}
\affiliation{Department of Astronomy and Astrophysics, University of California, San Diego, 9500 Gilman Dr, La Jolla, CA 92093, USA}

\author[0000-0002-8958-0683]{Fei Dai \texorpdfstring{(\cjk{戴飞})}{}}
\affiliation{Institute for Astronomy, University of Hawai`i, 2680 Woodlawn Drive, Honolulu, HI 96822, USA}

\author[0009-0009-6098-296X]{Bo Peng \texorpdfstring{(\cjk{彭博})}{}}
\affiliation{Department of Earth and Planetary Sciences, Stanford University, Stanford, CA 94305-2115, USA}

\author[0000-0003-3980-7808]{Yao Tang \texorpdfstring{(\cjk{唐尧})}{}}
\affiliation{Department of Astronomy and Astrophysics, University of California, Santa Cruz, 1156 High Street, Santa Cruz, CA 95064, USA}

\author[0000-0001-6974-6714]{Andrea Zorzi}
\affiliation{Department of Earth and Planetary Sciences, Stanford University, Stanford, CA 94305-2115, USA}

\author[0000-0002-0139-4756]{Michelle Hill}
\affiliation{Department of Earth and Planetary Sciences, Stanford University, Stanford, CA 94305-2115, USA}

\author[0000-0003-4241-7413]{Megan Weiner Mansfield}
\affiliation{Department of Astronomy, University of Maryland, College Park, MD 20742, USA}



\begin{abstract}
\noindent Recent JWST observations challenge the traditional \lq cosmic shoreline\rq{} from both sides, revealing thick volatile atmospheres on the hottest close-in \lq lava worlds,\rq{} where irradiation should drive the most extreme escape, and bare rocky surfaces on cooler terrestrial planets around M dwarfs, where atmospheres would be expected to survive. Using a coupled atmosphere-interior evolution model, we show that atmosphere retention is governed not by a single escape boundary but by two: a hot, outgassing-regulated \lq cosmic sandbar\rq{} and a cooler, escape-regulated \lq cosmic shoreline,\rq{} separated by an \lq airless valley\rq{} that may mark a graveyard of stripped sub-Neptune cores. The sandbar arises because long-lived magma oceans, sustained further by tidal heating from secular eccentricity excitation in multi-planet systems, keep most volatiles dissolved and expose only a small atmospheric reservoir to escape, whereas cooler planets solidify, sequestering volatiles in the deep solid mantle while overexposing the rest to loss. This two-regime structure recasts the single cosmic shoreline as two boundaries set by distinct physics: outgassing and escape. We provide time-evolving fits for both boundaries across G, K, and M stellar types as a function of volatile inventory, planetary mass, age, and tidal heating. Lava worlds with thick atmospheres are unlikely around stars cooler than K-type unless sustained by extreme tidal and/or other interior heating. This framework links atmosphere survival from USP lava worlds to habitable zone planets, informing target selection and interpretation for TRAPPIST-1 and JWST DDT characterization.

\end{abstract}
 


\section{Introduction}\label{sec:intro}

\citet{zahnle_cosmic_2017} proposed the \lq cosmic shoreline\rq, a boundary separating airless rocky planets from those retaining secondary atmospheres, shaped primarily by XUV irradiation, but also influenced by impact erosion \citep{owen_atmospheric_2019, kegerreis_atmospheric_2020,zhang_atmospheric_2020,wyatt_susceptibility_2020,krissansen-totton_erosion_2024}. While the trend explains the solar system well, differences in irradiation histories and escape physics around M dwarfs and other cool stars, where high XUV flux and extended XUV lifetimes suppress retention, may substantially alter the expected location of the shoreline compared to the solar system calibration \citep{france_ultraviolet_2013,france_muscles_2016,pass_receding_2025,coy_population-level_2025,ih_rocky_2025}. Broadly speaking, the sensitivity of the empirical shoreline to escape-regulated effects (escape mechanics, stellar regime, stellar and planetary dynamo activity, etc.) and replenishment-regulated effects (volatile abundances, outgassing, volcanism, orbital dynamics, etc.), among others, is being actively investigated \citep{gunell_why_2018,gronoff_atmospheric_2020,kite_exoplanet_2020,colose_effects_2021,chin_role_2024,ji_cosmic_2025,van_looveren_habitable_2025,august_atmospheric_2026,chatterjee_novel_2026,nguyen_effect_2026}. Additionally, the observed scaling between stellar instellation $I$ and planetary escape velocity $v_{\mathrm{esc}}$, $I \propto v_{\mathrm{esc}}^{4}$ \citep{zahnle_cosmic_2017, berta-thompson_3d_2025, meni-gallardo_empirical_2025}, which sets the boundary for predictions for planets with and without atmospheres, remains in conflict with the $I \propto v_{\mathrm{esc}}^{3}$ dependence predicted by energy-limited hydrodynamic escape \citep{zahnle_cosmic_2017,wordsworth_atmospheres_2022}, a tension most apparent among close-in rocky planets observed with JWST and Spitzer.

\subsection{USPs and Close-in Rocky Planets Challenge the \lq Cosmic Shoreline\rq}\label{subsec:intro:challenge}

Beyond the power-law discrepancy, the shoreline is increasingly challenged by tentative detections of thick volatile atmospheres on some ultra-short period planets (USPs), with orbital periods $P_{\text{orb}}<1$ day, despite instellation levels that should drive extreme escape. Additionally, campaigns such as the Rocky Worlds Director’s Discretionary Time (DDT) program \citep{redfield_report_2024} and ongoing TRAPPIST-1 observations suggest the framework requires revision for cooler planets, especially around M-type stars, where XUV histories, flare environments, and spectral contamination complicate atmospheric interpretations \citep{vida_frequent_2017,dong_modeling_2018,lim_atmospheric_2023}.

Spitzer and JWST thermal emission observations of hot rocky planets in the USP and near-USP ($P_\text{orb}< 2$ day) regime reveal an apparent host star dependence, likely reflecting an underlying irradiation effect on planetary surface melting. M dwarf targets LHS 3844 b, GJ 1252 b, GJ 367 b, TOI-1468 b, and LTT 3780 b all show airless, blackbody-like daysides consistent with bare rock emission and credible XUV-driven loss \citep{kreidberg_absence_2019, crossfield_gj_2022,zhang_gj_2024,meier_valdes_hot_2025,allen_hot_2025}. K-dwarf hosts yield intermediate cases, where K2-141 b shows limited heat redistribution and at most a tenuous rock vapor atmosphere \citep{zieba_k2_2022}, and TOI-431 b shows a dayside brightness temperature of $\sim$1520 K against a $\sim$2400 K bare rock prediction, suggesting a high albedo surface or modest heat redistribution \citep{monaghan_low_2025}. However, observations of HD 3167 b seem to favor a volatile atmosphere \citep{coy_evidence_2026}. Around G-type hosts, 55 Cnc e and TOI-561 b have JWST spectra requiring substantial volatile-rich atmospheres despite extreme irradiation \citep{hu_secondary_2024,teske_thick_2025}.

For cooler temperate close-in rocky planets, which mostly orbit M dwarf hosts due to favorable signal-to-noise (SNR), observations reveal a mixed picture dominated by bare rocks with a handful of tentative detections. 
TRAPPIST-1 b and TRAPPIST-1 c show hot daysides and weak heat redistribution, with TRAPPIST-1 b most likely airless and TRAPPIST-1 c strongly disfavoring a thick $\text{CO}_2$-rich atmosphere, though a thin atmosphere remains possible. \citep{de_wit_combined_2016, de_wit_atmospheric_2018, ducrot_0845_2018, greene_thermal_2023, zieba_no_2023, radica_promise_2025, gillon_no_2026}. Thick atmospheres are also unlikely on TRAPPIST-1 d, but observations cannot exclude a very thin atmosphere or one obscured by aerosols or clouds \citep{piaulet_ghorayeb_strict_2025}. Recent JWST/NIRSpec observations of TRAPPIST-1 e permit either a bare surface or an atmosphere with a high mean molecular weight, possibly N$_2$-rich atmosphere with trace CO$_2$/CH$_4$, but stellar contamination prevents a definitive interpretation \citep{espinoza_jwst_tst_2025, glidden_jwst-tst_2025}.  GJ 486 b and GJ 1132 b similarly favor bare rock interpretations \citep{moran_high_2023, may_double_2023, weiner_mansfield_no_2024, xue_jwst_2024, bello-arufe_evidence_2025}. Among planets with tentative atmospheric signals, LHS 1478 b remains unresolved due to systematics \citep{august_hot_2025}, L 98-59 b suggests a volcanically sustained $\text{SO}_2$-rich atmosphere \citep{bello-arufe_evidence_2025}, and TOI-270 b suggests an $\text{H}_2\text{O}$ steam atmosphere \citep{coulombe_possible_2025}. Most notably, LHS 1140 b shows tentative evidence for a possible $\text{N}_2$-rich secondary atmosphere with $\text{H}_2\text{O}$ and $\text{CO}_2$, potentially retained after the escape of any primordial H/He envelope \citep{cadieux_transmission_2024, damiano_lhs_2024, cherubim_helium_2026}. With further constraints on its atmospheric composition and surface pressure, LHS 1140 b may prove to be among the first temperate rocky M dwarf planets with an atmosphere on the retaining side of the cosmic shoreline. Together, these results suggest that stellar-planetary parameters and outgassing-escape dynamics blur the cosmic shoreline into a fuzzy transition.

In this paper, we generalize and extend the outgassing-escape framework of \citet{nguyen_effect_2026}, showing that the cosmic shoreline is better described as two boundaries, a hot cosmic sandbar and a cooler cosmic shoreline, separated by an airless valley reflecting dominant outgassing-regulated and escape-regulated regimes across a range of stellar and planetary properties. The sandbar arises because long-lived magma oceans can keep volatiles dissolved, exposing only a small atmospheric window to escape, whereas cooler planets solidify and force a larger fraction of their interior reservoir into the atmosphere, where it becomes vulnerable to erosion. As a result, close-in lava planets can sustain high present-day escape rates while retaining an observable atmosphere, cooler solidifying planets lose their atmospheres even as much of their volatile inventory is outgassed or sequestered into the mantle as trapped melt, and temperate planets beyond the shoreline can retain outgassed secondary atmospheres. This framework spans the emerging observational sequence, from lava worlds with possible thick volatile or silicate vapor atmospheres (55 Cnc e, TOI-561 b, HD 3167 b, TOI-431 b, K2-141 b), through apparently airless planets in the valley (LHS 3844 b, GJ 1132 b, TRAPPIST-1 b and c), to temperate atmosphere candidates such as LHS 1140 b. Section \ref{sec:methods} describes updates to our interior-atmosphere and escape model, Section \ref{sec:results} presents atmospheric outcomes across close-in rocky planets, and Section \ref{sec:discussion} addresses model limitations and implications for future rocky planet characterization of JWST DDT targets and target selection for HWO and Roman. 


\section{Hydrodynamic Escape and Interior-Atmosphere Model}\label{sec:methods}

In this section, we briefly describe updates to the XUV-driven hydrodynamic escape model and interior-atmosphere and climate model of \citet{schaefer_predictions_2016} and \citet{nguyen_effect_2026}. Section \ref{sec:methods:escape} outlines updates to the escape model, including a revised XUV and stellar luminosity prescription, and the resulting escape flux model for XUV-driven dissociation and atomic hydrodynamic loss. Section \ref{sec:methods:interior-atmosphere} then extends the coupled interior-atmosphere evolution framework to cover both the magma ocean stage and the post-solidification stage, where magma ocean outgassing shutoff occurs, with Section \ref{sec:methods:climate} consolidating its predicted effects on planetary climate and on the observational signatures of dayside–nightside heat redistribution. Section \ref{sec:methods:param} briefly provides an overview of the model grids simulated.

\subsection{XUV-driven Atomic Photodissociation and Loss in the Hydrodynamic Regime}\label{sec:methods:escape}

The stellar luminosity and XUV fraction evolution set the conditions for planetary atmospheric escape, governing both the energy available for hydrodynamic outflow and the photodissociation of outgassed volatiles into escaping atomic species. Here we update the stellar luminosity evolution and XUV model of \citet{nguyen_effect_2026} to extend to ultracool M (TRAPPIST-1), M (GJ 1132), K (TOI-500), and G (Sun-like) stellar types by incorporating updated XUV saturation and decay timescales, as well as bolometric luminosity tracks that reflect recent empirical constraints. In \citet{nguyen_effect_2026}, we adopted the stellar model of \citet{baraffe_new_2015} and \citet{ribas_evolution_2005}, in which the XUV luminosity ($L_{\mathrm{XUV}}$) is given by,

\begin{equation}\label{eqn:XUV_evol}
L_{\mathrm{XUV}}=
\begin{cases}
L_{\mathrm{bol}}\,f_0, & t<t_{\mathrm{sat}},\\[6pt]
L_{\mathrm{bol}}\,f_0\left(\dfrac{t}{t_{\mathrm{sat}}}\right)^{\beta_{\mathrm{XUV}}},
& t\ge t_{\mathrm{sat}},
\end{cases}
\end{equation}

\noindent where $f_0$ is the XUV saturation fraction, $t_{\mathrm{sat}}$ is the saturation timescale, and $\beta_{\mathrm{XUV}}$ is the post-saturation decay index. In this work, we adopt fiducial values from a mixture of published X-ray and XUV activity–age scalings and commonly used implementations. For each spectral class, we select a single representative sub-type as a proxy for the class as a whole: TRAPPIST-1 (M8) for ultra-cool M dwarfs \citep{birky_improved_2021}, an M4 dwarf for M stars more broadly \citep{schaefer_predictions_2016, pass_receding_2025}, a K4 for K stars \citep{ribas_evolution_2005, jackson_coronal_2012}, and a Sun-like G2 for G stars \citep{ribas_evolution_2005, nguyen_effect_2026}, parameterized as:

\begin{equation}\label{eqn:xuv_params}
\begin{aligned}
f_0 &=
\begin{cases}
9.33\times10^{-4}, & \text{TRAPPIST-1},\\
1.0\times10^{-3}, & \text{M star},\\
1.0\times10^{-3}, & \text{K star},\\
1.0\times10^{-3}, & \text{G star},
\end{cases}
\\[8pt]
t_{\mathrm{sat}} &=
\begin{cases}
3.14\,\mathrm{Gyr}, & \text{TRAPPIST-1},\\
1.00\,\mathrm{Gyr}, & \text{M star},\\
0.10\,\mathrm{Gyr}, & \text{K star},\\
0.10\,\mathrm{Gyr}, & \text{G star},
\end{cases}
\\[8pt]
\beta_{\mathrm{XUV}} &=
\begin{cases}
-1.17, & \text{TRAPPIST-1},\\
-1.23, & \text{M star},\\
-1.23, & \text{K star},\\
-1.23, & \text{G star}.
\end{cases}
\end{aligned}
\end{equation}

\noindent The stellar parameterizations above set the time-dependent XUV evolution that drives hydrodynamic atmospheric escape. We note that stellar type enters our escape model only through these bolometric luminosity tracks and XUV parameterizations. We do not employ wavelength-resolved stellar spectra, and we discuss the resulting spectral energy distribution caveats in Section~\ref{sec:discussion}. In this paper, we assume that under extreme XUV irradiation, volatile molecules dissociate into their atomic components in the exobase. This complete dissociation is an efficient-supply limit rather than a photochemical calculation, with dissociation altitudes and escape-level conditions ($T_{\mathrm{esc}}$, $P_{\mathrm{esc}}$) following Nguyen et al. (2026) and Schaefer et al. (2016). The resulting fluxes are therefore upper limits for a given atomic supply, which is in turn reduced for heavier species by their less efficient delivery to the escape region.

The binary diffusion coefficient $b_{ij}$ controls how effectively an escaping hydrogen wind entrains and removes heavier species, which we have previously adopted from the empirical fits of \citet{zahnle_mass_1986}. In this work, we instead compute $b_{ij}$ using an analytical Lennard-Jones Chapman-Enskog framework under a closed-shell assumption with ideal gas conversion, enabling direct calculation of diffusion interactions from the kinetic diameter $\sigma_{\text{LJ}, i}$, and potential energy well-depth $\epsilon_{LJ}/k_B$, tabulated in Table \ref{tab:LJ_diffusion_values}. The full Chapman-Enskog derivation of the resulting binary diffusion parameters $b_{\text{H},i}(T_\text{esc})$, together with their fitted power-law forms, is provided in Appendix~\ref{app:diffusion}.

Applying the updated binary diffusion coefficient $b_{\text{H},i}$ to the framework of \citet{nguyen_effect_2026}, we compute the reference hydrogen flux $\Phi_\text{H}^{\mathrm{ref}}$, crossover mass after calibration $\mu_{c_i}$, and calibrated crossover mass $\mu_{c_i}^{\mathrm{ref}}$ for each secondary species $i$ following their Equations (10), (8), and (7) of \citet{nguyen_effect_2026}. The resulting escape fluxes of atomic O and C ($\mathrm{m}^{-2}\ \mathrm{s}^{-1}$) are then given as:

\begin{equation}\label{eqn:carbon_escape}
\begin{aligned}
\Phi_{H,C}
&=\Phi_H^{\mathrm{ref}}
\left(\dfrac{\mu_{c_C}}{\mu_{c_C}^{\mathrm{ref}}}\right),\\
\Phi_C
&=
\begin{cases}
\Phi_{H,C}\dfrac{X_C}{X_H}\dfrac{\mu_{c_C}-\mu_C}{\mu_{c_C}-\mu_H},
&
\begin{aligned}
\Phi_{H,C}&\ge \Phi_{H,C}^{\mathrm{crit}},\\
P_{\mathrm{H_2O}}&>P_{\mathrm{CO_2}},
\end{aligned}
\\[10pt]
0,
&
\begin{aligned}
\Phi_{H,C}&< \Phi_{H,C}^{\mathrm{crit}},\\
\text{or}\quad P_{\mathrm{H_2O}}&\le P_{\mathrm{CO_2}},
\end{aligned}
\end{cases}
\end{aligned}
\end{equation}

\begin{equation}\label{eqn:oxygen_escape}
\begin{aligned}
\Phi_{H,O}
&=\Phi_H^{\mathrm{ref}}
\left(\dfrac{\mu_{c_O}}{\mu_{c_O}^{\mathrm{ref}}}\right),\\
\Phi_O
&=
\begin{cases}
\Phi_{H,O}\dfrac{X_O}{X_H}\dfrac{\mu_{c_O}-\mu_O}{\mu_{c_O}-\mu_H},
& \Phi_{H,O}\ge \Phi_{H,O}^{\mathrm{crit}},\\[8pt]
0, & \Phi_{H,O}<\Phi_{H,O}^{\mathrm{crit}},
\end{cases}
\end{aligned}
\end{equation}

\noindent We assume oxygen can be produced locally in the upper atmosphere by $\mathrm{H_2O}$ photolysis and is therefore co-located with the H wind in the escape region, making O loss limited primarily by hydrodynamic crossover and escape efficiency \citep{luger_extreme_2015, tian_history_2015,schaefer_predictions_2016}. By contrast, we treat carbon escape as supply-limited through the explicit pressure-ratio gate in Eq. \ref{eqn:carbon_escape}, so atomic C loss is permitted only under steam-dominated conditions, reflecting less efficient upward transport of heavier $\text{CO}_2$ to the dissociation and escape region, and the fact that $\text{CO}_2$-rich atmospheres can inhibit or complicate hydrodynamic loss through cooling and transport limits \citep{tian_thermal_2009, wordsworth_water_2013}. This carbon shutoff varies continuously with atmospheric state. The crossover drag factor $(\mu_{c_i}-\mu_i)/(\mu_{c_i}-\mu_H)$ drives entrained loss smoothly toward zero as the hydrogen flux weakens. At low steam pressures ($\lesssim 1$ bar), where upper-atmosphere photochemistry and heavy-species vertical transport are unresolved and may lose efficiency \citep{hunten_escape_1973,tian_thermal_2009,hu_photochemistry_2012,gronoff_atmospheric_2020,modirrousta-galian_diffusion_2024}, we further scale the carbon flux by a hyperbolic-tangent supply taper in $P_{\mathrm{H_2O}}$ that gradually falls from unity to zero below this threshold. Carbon escape therefore declines continuously as the steam reservoir is depleted toward exhaustion.

\subsection{Interior-Atmosphere Model}\label{sec:methods:interior-atmosphere}

Building on the atomic hydrodynamic escape model of Section \ref{sec:methods:escape}, in this Section we expand the framework of \citet{nguyen_effect_2026} to model thermal and volatile evolution from the magma ocean through the post-magma ocean stage, transitioning once surface temperatures fall below the melting temperature of peridotite (1420 K) and the mantle-averaged melt fraction drops below $\phi_{\mathrm{melt}}^{\mathrm{crit}}=0.4$, after which the model switches to a solid-mantle thermal evolution with minor volcanic outgassing. We also update the heat redistribution parameterization of \citet{koll_scaling_2022} to account for multi-component volatile atmospheres and their effect on the bulk atmospheric optical depth, enabling predictions of dayside-nightside heat redistribution within our interior-atmosphere evolution model. To map the simulation grid from planetary mass to radius, we adopt the empirical rocky planet mass-radius relation $R_p = R_\oplus (M_p/M_\oplus)^{0.27}$ for planets in the range 0.1--10 $M_\oplus$ following the generalized rocky planet scaling of \citet{valencia_internal_2006} and \citet{muller_mass-radius_2024}. We adopt a single Earth-composition scaling so that shifts in the shoreline and sandbar isolate the escape and interior-atmosphere physics rather than assumed bulk composition. Iron-enriched (Mercury-like) or volatile-rich compositions, or the thermally inflated radius of a molten mantle, would modestly shift $v_{\mathrm{esc}}$ and surface gravity \citep{zeng_growth_2019, dorn_hidden_2021, unterborn_nominal_2023}, which we expect to minimally displace individual planets relative to the fitted boundaries.

For the thermal evolution, we adopt the mantle ($T_\mathrm{mantle}$) and surface temperature ($T_\mathrm{surf}$) evolution equations of \citet{nguyen_effect_2026} (their equations (26) and (28), respectively), but reparameterize the solidification front as a function of the globally-averaged melt fraction $\phi_{\mathrm{melt}}$ beyond a permanent magma ocean stage:

\begin{equation}\label{eqn:solidification_r}
\frac{dr_s}{dt}
=
\begin{cases}
\frac{dT_{\mathrm{mantle}}}{dt}\,
\frac{c_{p,m}\left(b\alpha-a\rho_m c_{p,m}\right)}
{g\left(a\rho_m c_{p,m}-\alpha T_{\mathrm{mantle}}\right)^2},
& \phi_{\mathrm{melt}}\ge \phi_{\mathrm{melt}}^{\mathrm{crit}},\\[12pt]
\displaystyle
0,
& \phi_{\mathrm{melt}}<\phi_{\mathrm{melt}}^{\mathrm{crit}},
\end{cases}
\end{equation}

\noindent where in the magma ocean stage ($\phi_{\mathrm{melt}} \geq \phi_{\mathrm{melt}}^{\mathrm{crit}}$), $r_s$ is the radius of the solidification front advancing bottom-up from the core-mantle boundary (CMB), and in the post-magma ocean stage ($\phi_{\mathrm{melt}} < \phi_{\mathrm{melt}}^{\mathrm{crit}}$), the rheological front has reached the surface and is stationary. Here $a$ and $b$ are the linear fit coefficients of the solidus temperature profile $T_{\mathrm{sol}}(P)$ from \citet{nguyen_effect_2026} equation (18), $\alpha$ is the thermal expansion coefficient, $\rho_m$ is the mantle density, $c_{p,m}$ is the mantle heat capacity, and $g$ is the surface gravitational acceleration. The resulting thermal evolution of the mantle ($T_{\mathrm{mantle}}$) and surface ($T_{\mathrm{surf}}$) temperatures therefore depend on the progression of the solidification front $r_s$ and the transition in mantle viscosity from liquid-like to solid-body whole-mantle convection, as outlined in \citet{nguyen_effect_2026}.

Volatile evolution is solved via mass balance between the solid mantle, magma ocean, and atmosphere using the solubility partition equations of \citet{nguyen_effect_2026} equations (29-41). We do not include evolution of the core, though the consequences of core volatile sequestration and metal–silicate differentiation for the mantle volatile budget are discussed in Section \ref{sec:discussion}. Equation (\ref{eqn:volatile_evol_sol}) describes mantle sequestration during crystallization, parameterized by the tracked volatile species $i$, its solid-liquid partition coefficient $k_i$, and its fractional mass in melt $F_i^{\mathrm{liq}}$, active only in the magma ocean stage. During this stage, oxygen is specifically governed by redox buffering through $\mathrm{FeO}_{1.5}$ \citep{hirschmann_magma_2012}, the $\mathrm{Fe}^{3+}$-bearing oxide component of the melt (equivalent to $\mathrm{Fe}_{2}\mathrm{O}_{3}$ written per mole of Fe). Equation (\ref{eqn:volatile_evol_liq_atm}) describes the coupled atmosphere-magma ocean system during the magma ocean stage, reducing to atmospheric hydrodynamic loss alone with negligible volcanic outgassing in the post-solidification stage, given as,

\enlargethispage{\baselineskip}
\begin{equation}\label{eqn:volatile_evol_sol}
\begin{aligned}
\frac{dM_i^{\mathrm{sol}}}{dt}
&=
\frac{dr_s}{dt}\;
k_i\,F_i^{\mathrm{liq}}\;4\pi\,\rho_m\,r_s^{2},\\
\frac{dM_{\mathrm{O}}^{\mathrm{sol}}}{dt}
&=
\frac{dr_s}{dt}\,F_{\mathrm{FeO}_{1.5}}^{\mathrm{liq}}\;4\pi\,\rho_m\,r_s^{2}\;
\frac{\mu_{\mathrm{O}}}{2\,\mu_{\mathrm{FeO}_{1.5}}}.
\end{aligned}
\end{equation}

\begin{equation}\label{eqn:volatile_evol_liq_atm}
\begin{aligned}
\frac{dM_{\mathrm{H_2O}}^{\mathrm{liq+atm}}}{dt}
&= -\frac{dM_{\mathrm{H_2O}}^{\mathrm{mantle,volc.}}}{dt}
\\
&\quad -\frac{dM_{\mathrm{H_2O}}^{\mathrm{sol,MO}}}{dt}
      -4\pi R_p^2 \Phi_H \frac{\mu_{\mathrm{H_2O}}}{2\mu_H},
\\
\frac{dM_{\mathrm{CO_2}}^{\mathrm{liq+atm}}}{dt}
&= -\frac{dM_{\mathrm{CO_2}}^{\mathrm{mantle,volc.}}}{dt}
\\
&\quad -\frac{dM_{\mathrm{CO_2}}^{\mathrm{sol,MO}}}{dt}
      -4\pi R_p^2 \Phi_C \mu_{\mathrm{CO_2}},
\\
\frac{dM_{\mathrm{O}}^{\mathrm{liq+atm}}}{dt}
&= -\frac{dM_{\mathrm{O}}^{\mathrm{sol,MO}}}{dt}
\\
&\quad +4\pi R_p^2\Bigg(
      \Phi_H\frac{\mu_O}{2\mu_H}
      +2\Phi_C\mu_O
      -\Phi_O
   \Bigg),
\end{aligned}
\end{equation}

\noindent where for the magma ocean stage, $dM_{i}^{\mathrm{mantle,volc.}}=0$ as solid-mantle volcanism cannot occur prior to crustal solidification, and $M_{i}^{\mathrm{sol,MO}}$ tracks volatiles sequestered during mantle crystallization. For the post-magma ocean stage, solidification reaches the planetary radius ($r_s = R_p$), so $dM_{i}^{\mathrm{sol,MO}}=0$, and solid-mantle volcanic outgassing is set to $dM_{i}^{\mathrm{mantle,volc.}} \approx 0$, as post-solidification volcanic outgassing is neglected due to uncertainties in population-scale planetary tectonics and volcanism \citep{foley_carbon_2018, kite_exoplanet_2020, quick_forecasting_2020, reinhold_ignan_2025, hill_smaller_2026}.

\subsection{Climate and Heat Redistribution Model}\label{sec:methods:climate}

The coupled interior-atmosphere model of Section \ref{sec:methods:interior-atmosphere} yields atmospheric compositions as a function of time, orbital distance, and planet mass. In this Section, we feed those evolving atmospheric states into a climate and heat redistribution model to estimate the dayside surface temperature, enabling direct comparison with brightness temperatures inferred from secondary-eclipse observations.

To calculate heat redistribution for mixed-composition atmospheres, we parameterize the redistribution efficiency $\varepsilon$ as a function of whole-atmosphere longwave optical depth $\tau_{\mathrm{LW}}$, adopting the analytic framework of \citet{koll_scaling_2022} and \citet{lin_persistent_2026}. We quantify redistribution using the dayside temperature ratio \citep{weiner_mansfield_no_2024, xue_jwst_2024, lin_persistent_2026},

\begin{equation}\label{eqn:R_ratio}
\mathcal{R} \equiv \frac{T_{\mathrm{day}}}{T_{\mathrm{day,max}}}
= \left(1-\frac{5}{8}\varepsilon\right)^{1/4},
\end{equation}

\noindent where $T_{\mathrm{day}}$ is the dayside emission temperature, $T_{\mathrm{day,max}}$ is the maximum dayside temperature in the zero redistribution limit, and $\varepsilon$ is the heat redistribution efficiency following \citet{lin_persistent_2026} equations (3), (4), and (16). In the \citet{koll_scaling_2022} convention, the full redistribution limit corresponds to $\mathcal{R} = (3/8)^{1/4} = 0.783$, whereas the zero redistribution limit for airless blackbodies is $\mathcal{R} = 1$, though the baseline thick atmosphere will correspond to weakening heat redistribution as planetary irradiation temperature $T_\mathrm{irr}$ (\citet{lin_persistent_2026} equation (2)) increases. Whether Bond albedo $A_\mathrm{B}$ appears explicitly in the ratio $\mathcal{R}$ depends on the adopted definition of $T_{\mathrm{day,max}}$. $A_\mathrm{B}$ cancels out if both $T_{\mathrm{day}}$ and $T_{\mathrm{day,max}}$ are referenced to the same absorbed stellar flux, which we adopt here as a simplifying approximation across the modeled rocky-planet parameter space and which is broadly consistent with low albedos expected for irradiated bare rocks \citep{coy_population-level_2025}. However, $A_\mathrm{B}$ is retained if $T_{\mathrm{day,max}}$ is defined relative to a zero-albedo reference temperature \citep{xue_jwst_2024}, in which case $\mathcal{R}\propto(1-A_\mathrm{B})^{1/4}$. Additionally, we calculate $\tau_{\mathrm{LW}}$ from \citet{nguyen_effect_2026} equation (4), following the atmospheric composition predicted by our interior-atmosphere model, enabling composition-dependent heat redistribution across evolving volatile atmospheres.

\subsection{Model Parameter Grid}\label{sec:methods:param}

We run roughly 50 by 50 model grids in planet mass and semi-major axis as a function of stellar type, initial volatile content, and eccentricity. Initial H$_2$O and CO$_2$ inventories are assigned as fractions of the total planet mass. Our nominal case adopts $F_{\rm H_2O}=10^{-3}$, $F_{\rm CO_2}=10^{-4}$, and $e=0$, chosen such that Venus and Earth fall on the atmosphere-retaining side of the resulting G star shoreline, as a sanity check that the model recovers familiar outcomes for known cases. The volatile-poor case adopts $F_{\rm H_2O}=10^{-3}$, $F_{\rm CO_2}=10^{-5}$, and $e=0$, while the non-zero eccentricity case uses the nominal volatile budget with $e=0.01$. For eccentric cases, tidal heating is calculated assuming constant eccentricity following \citet{nguyen_effect_2026}. Because close-in rocky planets are expected to circularize unless eccentricity is continually re-excited \citep{bolmont_tidal_2013, ferraz-mello_tidal_2025}, our $e=0.01$ grid should be interpreted as an upper-limit tidal heating case, with more realistic intermediate eccentricities ($e=0.001$ to $e=0.005$) expected to span the parameter space between the $e=0$ and $e=0.01$ grids. We emphasize that the $e=0.01$ case is not intended to describe isolated close-in planets, for which circularization timescales are short ($\lesssim$0.1--3 Myr, \citealt{ferraz-mello_tidal_2025}). It instead represents compact, high-multiplicity architectures in which secular interactions continually re-excite small but non-zero eccentricities over Gyr timescales, and should be read throughout as a sustained-forcing upper limit on tidal heating rather than an assumption of primordial eccentricity retention.



\begin{figure*}
\center
\includegraphics[width = 2.15\columnwidth]{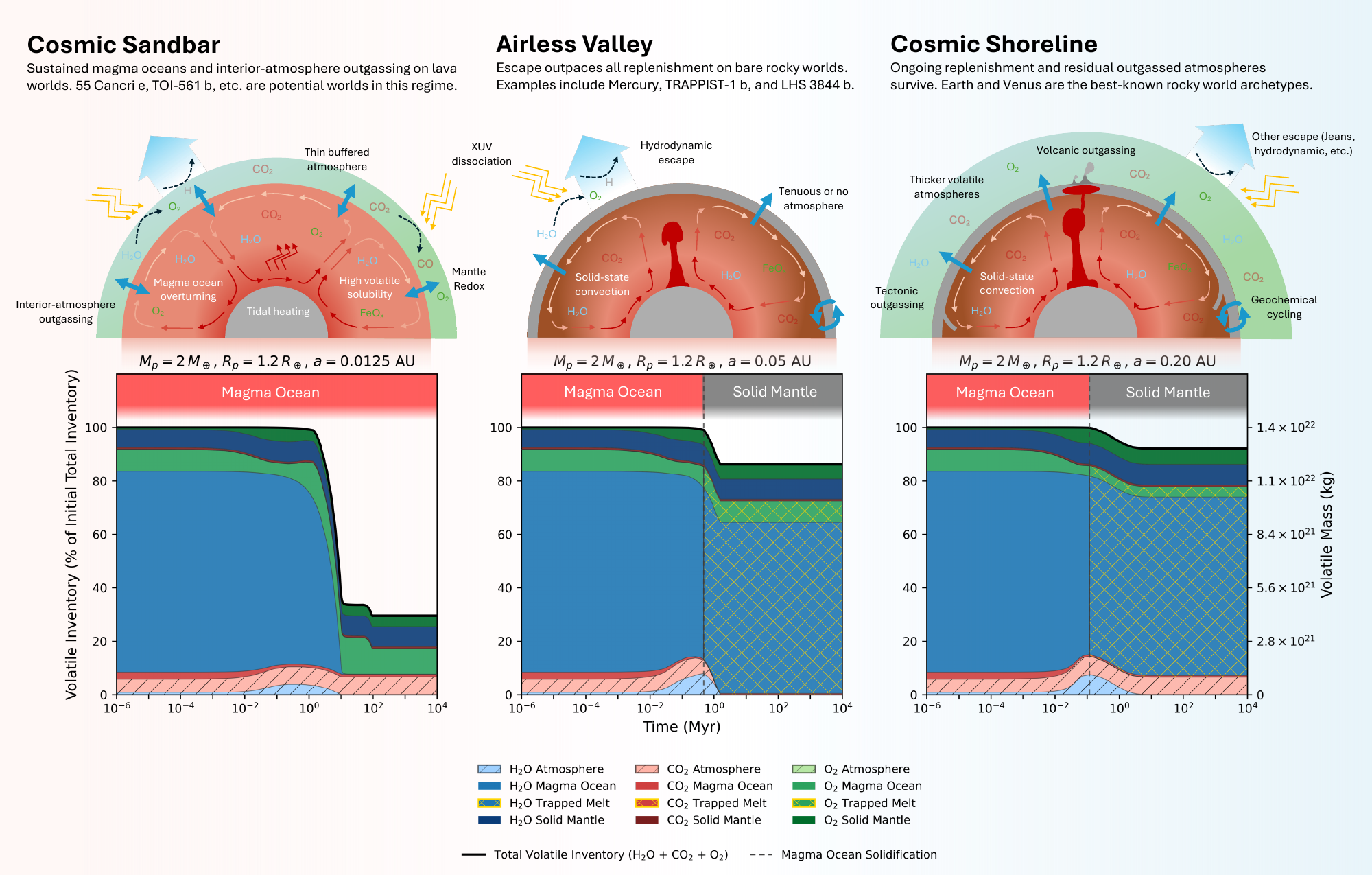}
\caption{Schematic of the three atmospheric retention and loss regimes for close-in rocky planets. Illustrated diagrams appear above stacked area charts tracking the volatile reservoir evolution of representative fiducial planets (2 $ M_{\oplus}$, 1.2 $R_{\oplus}$, G star host, $F_{\mathrm{H_2O}} = 10^{-3}$, $F_{\mathrm{CO_2}} = 10^{-4}$, $e = 0$). Colored regions show retained volatile inventories in the atmosphere, magma ocean, trapped melt, and solid mantle as percentages of the initial total volatile inventory, with equivalent kg on the right axis. The black curve shows the total retained H$_2$O, CO$_2$, and O$_2$-equivalent free oxygen inventory. Atmospheric O is tracked as molecular O$_2$, while magma ocean and mantle O are bound as ferric oxide, including the initial primordial concentration. The yellow-hatched band marks the uncertainty of residual melt sealed at the rheological transition. We assume no post-solidification outgassing, so this uncertainty band represents the possible lower and upper range of total volatile retention in the interior reservoir, since compaction, geochemical cycling, tectonics, volcanism, late-stage crystallization, and other residual outgassing could further drain the interior and return volatiles to the atmosphere. \citep{elkins-tanton_linked_2008, hirschmann_magma_2012, hamano_emergence_2013, lebrun_thermal_2013, schaefer_predictions_2016, hier_majumder_origin_2017, lichtenberg_coupled_2026}. \textbf{Left:} In the fiducial outgassing-regulated sandbar example ($a = 0.0125$ AU), a sustained magma ocean keeps most volatiles dissolved in the melt and exposes only a thin atmosphere to escape. \textbf{Middle:} In the fiducial airless valley example ($a = 0.05$ AU), magma ocean solidification partitions volatiles between the atmosphere and sequestered interior, after which atmospheric escape outpaces replenishment. \textbf{Right:} For a representative planet beyond the escape-regulated shoreline ($a = 0.20$ AU), outgassed atmospheres survive over Gyr timescales.}
\label{fig:art_schematic}
\end{figure*}

\section{Results}\label{sec:results}

In this section, we present results for an atmosphere–interior exchange model with coupled atmospheric escape, as described in the previous section. We show that this model reproduces a boundary separating planets with and without atmospheres, similar to the traditional cosmic shoreline of \citet{zahnle_cosmic_2017}, but that this boundary is dependent on the planet's volatile budget, age, stellar host type, and tidal heating. We define this boundary at a surface pressure of 0.01 bar, approximately Mars's surface pressure, following the convention adopted in prior shoreline studies, such that planets below this threshold are typically classified as bare rocky bodies and those above as retaining an atmosphere. However, we note that this 0.01 bar boundary is used here only as a fiducial Mars-like benchmark for a tenuous residual atmosphere, rather than as a uniquely defined observational or optical boundary, which may vary with atmospheric composition, mean molecular weight, and cloud opacity. 

We also identify two new features in atmosphere-retention space. Around hotter stars, the most close-in planets retain atmospheres, but they are separated from the cooler atmosphere-bearing population by an intermediate region in which all atmospheres are lost. We dub the inner atmosphere-bearing region the \lq cosmic sandbar\rq{} and the intervening barren region the \lq airless valley,\rq{} bounded on its outer edge by the traditional cosmic shoreline. Figure~\ref{fig:art_schematic} summarizes the three regimes schematically, pairing each with the modeled time evolution of its atmosphere, magma ocean, and solid mantle volatile reservoirs for a fiducial planet. Because the simulated shorelines contain non-monotonic structure in mass-semi-major axis space, the plotted fits that we provide should be interpreted as simple empirical summaries of the transition region rather than precise physical boundaries at every location. We discuss both of these results in the following two subsections, along with comparisons to the observed small planet atmosphere detections and non-detections. 

\subsection{Dependence of An Escape-Regulated Shoreline on Volatile Budget and Stellar Type}\label{sec:shoreline-budget}
Figure \ref{fig:3x3_cosmic_shoreline} presents the evolution of atmospheric pressure against the action of atmospheric escape and atmosphere-mantle exchange across three stellar types (G star, M star, and TRAPPIST-1) under three sets of initial conditions varying volatile inventory and orbital eccentricity. Although we calculate evolution models for a nominal K star, these are not shown here for space and because they strongly resemble the results for the G star. We calculate tidal heating rates for a constant orbital eccentricity as in \citet{nguyen_effect_2026}, and here we consider only one nominal eccentricity value of 0.01. As defined in Section~\ref{sec:methods:param}, the nominal, volatile-poor, and non-zero eccentricity grids isolate the effects of volatile inventory and tidal heating on the shoreline morphology. We focus our parameter study on the edges of the shoreline set primarily by magma ocean outgassing, rather than the deeper interior or extended regions where outgassing from a solid mantle through volcanism may dominate, which we do not model here.

The numbered contours in Figure \ref{fig:3x3_cosmic_shoreline} show the total atmospheric pressure (bars) at the end of 5 Gyr of evolution, with colors indicating the major gas composition. Almost all atmospheres are dominated by CO$_2$, although small regions exterior to the cosmic shoreline are dominated by O$_2$, which is produced by loss of H from water vapor and inefficient drag of the associated O atoms. We overlay blue-colored fits showing the position of the cosmic shoreline at different ages for the planet. We identify the shoreline with the contour of atmospheric pressures of 0.01 bars, approximating Mars's atmosphere, which we fit with a scaling-polynomial given in Table \ref{tab:polyfit_au_mass}. 

For our nominal volatile case, the cosmic shoreline position for G-type stars is relatively stable, whereas there is more temporal evolution for the smaller nominal M star and TRAPPIST-1 star, most likely a consequence of faster planetary cooling and outgassing, as well as differing stellar luminosity and XUV evolution histories. Both of the cooler stars have cosmic shorelines at much shorter orbital periods, consistent with cooler stellar temperatures. We find that tidal heating does not play a role in setting the outer cosmic shoreline for any of the stellar types that we explore here. The greater inward shift of the traditional outer shoreline for smaller planets (approx. $<1M_\oplus$), especially for the volatile-poor cases, reflects faster initial outgassing from more rapid cooling, followed by more efficient atmospheric escape because of their lower escape velocities, especially during the young active stellar phase. Planets initially outgas mixed steam and $\mathrm{CO_2}$-rich atmospheres, but their subsequent evolution depends strongly on mass. XUV photolysis converts residual steam into H and O, while hydrodynamic escape removes these products with an efficiency set by gravity and the crossover mass. At lower masses, escape can remove H while also dragging heavier products such as O and C, so planets can pass through mixed steam/O$_2$ states before being stripped to bare rock. At higher masses, H is preferentially removed from photolyzed steam while O and C escape are less efficient, allowing O$_2$-rich atmospheres to accumulate. As $\mathrm{H_2O}$ is depleted, continued carbon escape becomes limited because $\mathrm{CO_2}$ is heavier and less efficiently supplied to the escaping region, while O escape can continue, driving the retained atmosphere toward a lower pressure $\mathrm{CO_2}$-dominated residual state. This mass-dependent fractionation produces the inward bend of the shoreline, separating readily stripped lower mass planets from higher mass planets that retain volatile atmospheres. In our models, this transition occurs near $\sim0.5$--$0.8M_\oplus$.

For the volatile-poor cases, the cosmic shorelines for all of the stellar types move further outward, and all also show considerably more temporal variability, with significant structure in mass-semi-major axis space at 10 and 100 Myr. This structure arises because volatile-poor planets pass near an atomic crossover mass-dependent hydrodynamic fractionation transition, where small changes in planet mass, orbital distance, and volatile supply determine whether planets evolve toward residual oxygen, carbon retention, or bare rock outcomes. This sensitivity is most apparent as a bump in the residual $\mathrm{H_2O}$-to-$\mathrm{O_2}$ transition between 10 and 100 Myr, hinting at more efficient abiotic $\mathrm{O_2}$ production from XUV photolysis on smaller planets. The variable rate of escape for different volatile budgets contributes to the `fuzziness' of the cosmic shoreline noted in other work \citep{berta-thompson_3d_2025}. 

To compare with previous cosmic shoreline models \citep{zahnle_cosmic_2017, berta-thompson_3d_2025, meni-gallardo_empirical_2025}, we convert the shoreline boundaries identified in Figure \ref{fig:3x3_cosmic_shoreline} into escape velocity--cumulative XUV fluence space by integrating Eq. \ref{eqn:XUV_evol} for each planet. The result is shown in the top panels of Figure \ref{fig:cosmic_shoreline}. Fits for this parameterization are given in Table \ref{tab:polyfit_IXUV_ves}. The traditional cosmic shoreline of \citet{zahnle_cosmic_2017} falls largely between the cosmic shoreline boundaries that we compute for G stars using the nominal and volatile-poor cases, although the slope is slightly steeper at larger planet masses. The Sun-like luminosity shoreline relationship from \citet{berta-thompson_3d_2025} and shoreline model of \citet{meni-gallardo_empirical_2025} are both considerably steeper than our relationships, which would produce testable differences in atmosphere retention for planets around mature stars in the intermediate region between our respective shoreline boundaries. 


\subsection{An Outgassing-Regulated Cosmic `Sandbar' and `Airless Valley' for Hotter Stars and Tidally-Heated Planets}\label{sec:sandbar}
In Figure \ref{fig:3x3_cosmic_shoreline}, we see that for the nominal volatile budget the G star case produces substantial remaining atmospheres on super-Earths at short orbital distances even out to 5 Gyr. For the volatile-poor case, these atmospheres only survive for the first 10 Myrs before being eroded away. This region of close-in atmospheres, which we dub the `cosmic sandbar,' is separated from the atmosphere bounded by the cosmic shoreline by a region we dub the `airless valley'. We highlight the location of the sandbar with a red polynomial fit in each of the panels. We present polynomial best-fit scaling relations in mass-orbital distance space for the time-evolving cosmic sandbars as a function of planetary parameters in Table \ref{tab:polyfit_au_mass}. 

The sandbar is produced by a continuous magma ocean, which outgasses only minimally because lava worlds can remain substantially molten in prolonged magma ocean states \citep{nicholls_self-limited_2025, nguyen_effect_2026}. For these planets, even a tenuous atmosphere can be coupled to a much larger dissolved volatile reservoir in the melt. Using the H$_2$O partitioning relations of \citet{nguyen_effect_2026}, adapted from \citet{papale_modeling_1997}, an example 0.01 bar H$_2$O atmosphere on a rocky Earth-mass planet corresponds to only $\sim 5\times10^{16}\text{ kg}$ of atmospheric H$_2$O, but to a dissolved magma ocean reservoir of $\sim 2\times10^{19}\text{ kg}$, for a fully molten mantle magma ocean, about 400 times more volatile-rich in the interior. This buffering is not specific to the example above. Because the dissolved volatile mass $M_{i}^{\mathrm{liq}}$ scales with melt mass $M_{\mathrm{liq}}$ while the dissolved concentration $C_i$ can rise sublinearly with volatile partial pressure $P_i$ \citep{papale_modeling_1997}, the dissolved-to-atmospheric ratio for a volatile species $i$ is schematically

\begin{equation}\label{eqn:volatile_buffering_ex}
\frac{M_{i}^{\mathrm{liq}}}{M_{i}^{\mathrm{atm}}}
\simeq
\frac{C_{i}(P_i,T)\,M_{\mathrm{liq}}}
{4\pi R_p^2 P_{i}/g}
\propto
\frac{g\,M_{\mathrm{liq}}}{4\pi R_p^2}
P_{i}^{\alpha_i-1},
\end{equation}

\noindent where $C_{i}\propto P_{i}^{\alpha_i}$, with $\alpha_{\mathrm{H_2O}}<1$ for sublinear H$_2$O solubility. For such species, the ratio increases with connected melt mass, planetary mass, and melt fraction. For a sustained magma ocean, it also rises as escape lowers the atmospheric pressure, since the atmospheric reservoir falls nearly linearly with pressure while the dissolved reservoir falls more slowly. Escape therefore leaves behind a progressively thinner but proportionally more melt-buffered atmosphere in long-lived or permanent magma oceans, a buffering that weakens once crystallization shrinks the connected melt reservoir and expels volatiles into the atmosphere during magma ocean solidification. Figure~\ref{fig:art_schematic} illustrates this partitioning across fiducial sandbar, airless valley, and shoreline planets. Escape can therefore act only on a small atmospheric window while the bulk of the volatile inventory remains buffered and shielded in the melt. Even if extreme hydrodynamic escape reduces a thicker atmosphere to a tenuous state near the 0.01 bar threshold, the system can still remain in this same solubility-buffered, supply-limited regime, such that only enough atmosphere is exposed at any given time to balance ongoing loss, while heavier species escape weakens once the H$_2$O-derived hydrogen wind falls below the critical drag threshold $\Phi_{\mathrm{H},i}^{\mathrm{crit}}$ in these thinner atmospheres. Compositional differences may further reinforce this self-limiting behavior. Because H$_2$O is far more soluble in silicate melt than CO$_2$, the thin atmosphere above a long-lived magma ocean may trend CO$_2$-dominated and hydrogen-poor where XUV-driven escape is too weak to draw the dissolved H$_2$O back out of the melt, sustaining only a weak H wind that inefficiently entrains heavier species \citep{lichtenberg_coupled_2026}. By contrast, super-Earths in the airless valley cool and begin to solidify, driving the system out of this solubility-buffered state. Because solid silicate stores volatiles far less efficiently than melt, crystallization expels much of the dissolved inventory into the atmosphere, exposing it to escape without further interior resupply, while locking the remainder into the escape-shielded solid mantle \citep{nguyen_effect_2026}. Solidification therefore enhances atmospheric erosion even as it sequesters part of the inventory.

For G stars, the nominal case produces a sandbar which recedes from 10 Myr to 5 Gyr. The volatile-poor case eliminates the sandbar by 5 Gyr due to atmospheric loss and volatile depletion. For cooler M stars, sandbars do not form for any modeled volatile reservoirs, as cooler equilibrium temperatures are insufficient to sustain magma oceans and the continued outgassing needed to offset atmospheric loss. However, tidal heating can promote magma ocean formation and longevity for all stars. For G stars, the non-zero eccentricity case broadly matches the nominal case, but tidal heating prolongs cooling timescales, expanding the sandbar to include rocky sub-Earth mass planets. For our nominal M star and TRAPPIST-1 case,  tidal heating prolongs magma ocean lifetimes to Gyr timescales, allowing the most close-in rocky planets to compete against XUV-driven loss and retain a tidally-enhanced volatile atmosphere.  

To summarize, sandbars and shorelines evolve as a function of both stellar and planetary parameters. Planetary mass governs initial volatile budgets, cooling timescales, and outgassing rates, ultimately shaping the balance between volatile escape and replenishment over the planetary lifetime. Stellar XUV history, mass-luminosity, and effective temperature further modulate this balance through their influence on tidal heating, equilibrium temperature, and cumulative XUV-driven atmospheric loss. These effects yield a non-monotonic dependence on planetary mass. Sub-Earth mass planets cool faster, driving outgassing earlier in the planetary lifetime when XUV flux is most conducive to dissociation and loss, while their smaller volatile reservoirs and lower gravity further limit atmospheric retention. Super-Earth mass planets are the most likely to retain substantial atmospheres, owing to larger volatile reservoirs, higher surface gravity, and prolonged cooling timescales that can outlast the stellar XUV saturation phase. The tension between the observed $I \propto v_{\rm esc}^{4}$ scaling and the energy-limited escape prediction of $I \propto v_{\rm esc}^{3}$ may be reconciled by interpreting the classical shoreline as two distinct atmospheric regimes: the outgassing-regulated sandbar and the escape-regulated shoreline.

\subsection{Comparison with Observed Atmospheric Detections and Non-Detections}\label{sec:observations}

\begin{figure*}
\center
\hspace*{-4mm}\includegraphics[width=1.9\columnwidth]{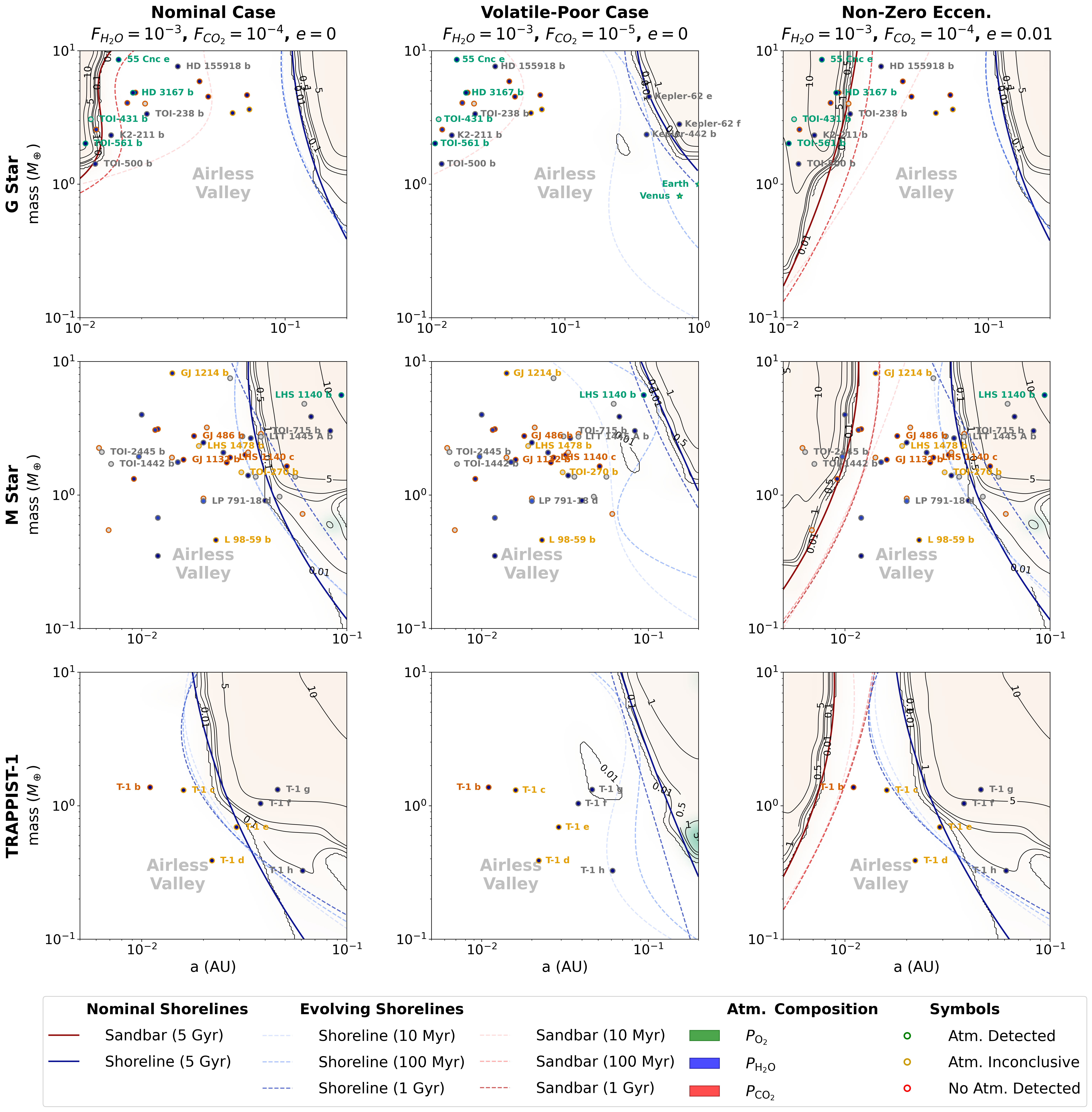}
\caption{A 50 by 50 grid of planetary mass ($M_\oplus$) and orbital distance (AU) showing contours of total atmospheric pressure after 5 Gyr, including pressures $>1$ bar, relevant for future JWST, HWO, and Roman detectability. Columns show the nominal volatile budget on circular orbits (left), reduced volatile budgets (middle), and the nominal budget with tidal heating at constant $e=0.01$ (right). The $e=0.01$ column is a sustained-forcing upper limit for compact, high-multiplicity systems in which secular excitation maintains non-zero eccentricity, as isolated short-period planets circularize within $\lesssim$0.1--3 Myr. Red and blue curves mark the outgassing-regulated cosmic sandbar and the escape-regulated cosmic shoreline, solid at 5 Gyr and dashed at 10 Myr, 100 Myr, and 1 Gyr. Contours are shaded by predicted atmospheric composition. Observed rocky exoplanets are overlaid by stellar type from the DDT survey, \citet{pass_receding_2025}, the NASA Exoplanet Archive, and the IAC ExoAtmospheres database. Best-fit relations are provided in Table~\ref{tab:polyfit_au_mass}. In some TRAPPIST-1 cases the fitted shoreline appears to move outward with time at low masses, reflecting the limits of low-order polynomial fits to a strongly curved boundary, so below $\sim0.3\,M_\oplus$ the fits should be read as approximate guides.}
\label{fig:3x3_cosmic_shoreline}
\end{figure*}

In Fig. \ref{fig:3x3_cosmic_shoreline}, we note that many of the small planets with atmosphere detections fall into the sandbar region of the tidally-heated K-G star case, including 55 Cnc e, TOI-431 b, TOI-561 b, and HD 3167 b. Several planets with strong indications that they do not have atmospheres, such as CoRoT-7b, also fall in this region, which may indicate that they either do not experience ongoing tidal heating or that they had lower initial volatile budgets. For the TRAPPIST-1 planets in the bottom row, TRAPPIST-1 b, and possibly TRAPPIST-1 c and d, are predicted to be likely bare rocks in the airless valley for the nominal volatile case, unless sustained by tidally-driven stochastic volcanism. TRAPPIST-1 e may harbor a thin, Mars-like atmosphere, while f and g may retain thicker atmospheres but are cold enough to risk nightside collapse without heat redistribution \citep{hu_role_2014, wordsworth_atmospheric_2015, august_atmospheric_2026}. TRAPPIST-1 h is unlikely to retain any atmosphere given its low mass and cold equilibrium temperature, and may resemble an airless icy world. In the volatile-poor case, reduced volatile reservoirs render all TRAPPIST-1 planets to be bare rocks.

\begin{figure*}
\center
\includegraphics[width = 2.15\columnwidth]{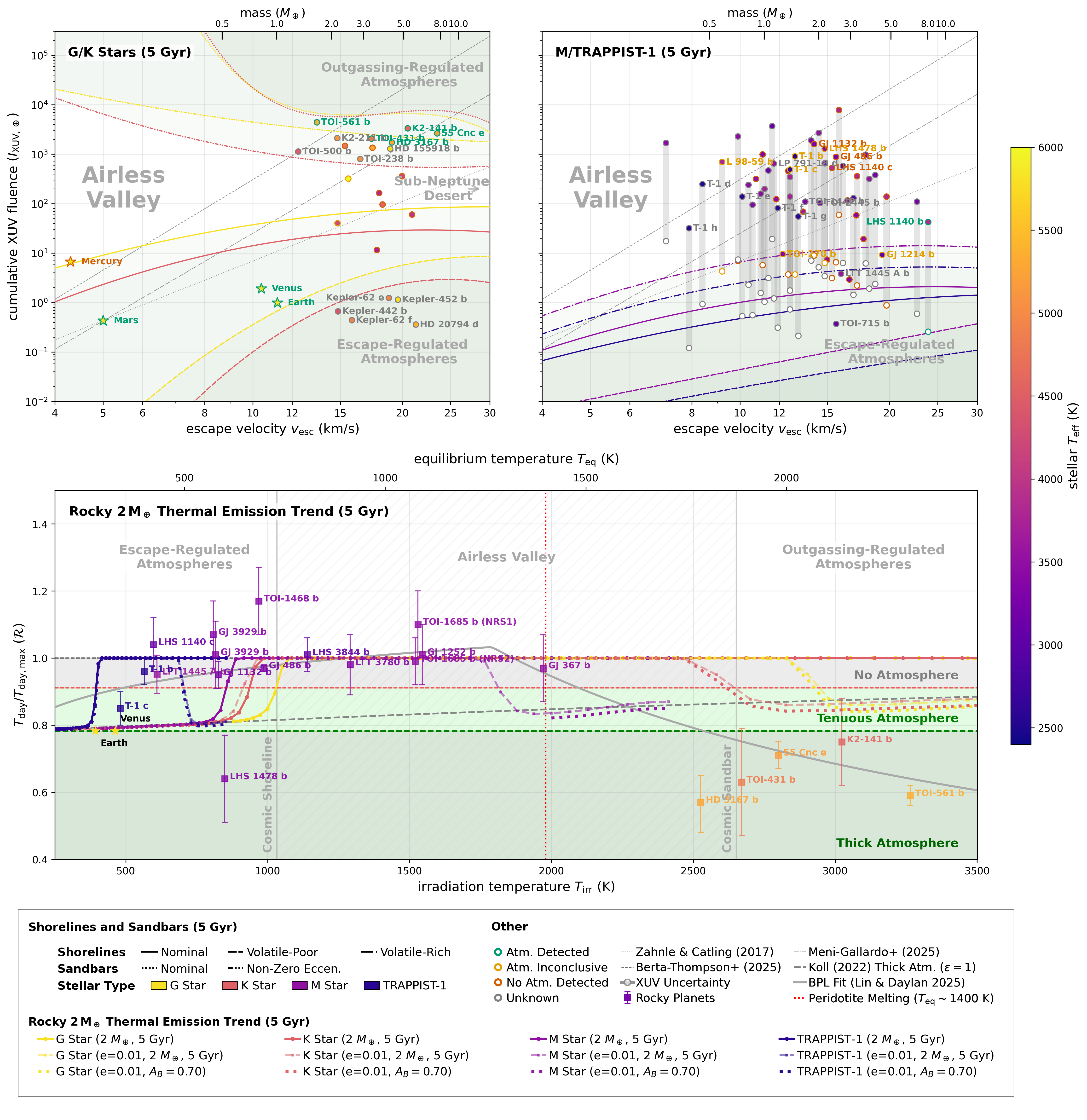}
\caption{Evolving shorelines and sandbars at 5 Gyr, including the cases of Figure \ref{fig:3x3_cosmic_shoreline} and a volatile-rich case ($F_{\rm H_2O}=10^{-2}$, $F_{\rm CO_2}=10^{-3}$, $e=0$) for M stars and TRAPPIST-1, motivated by possible water-rich formation. {\bf Top Left \& Right:} Cosmic sandbars and shorelines from Figure~\ref{fig:3x3_cosmic_shoreline} in escape velocity--XUV fluence space, with best-fit relations in Table~\ref{tab:polyfit_IXUV_ves}. We do not show the sandbar for M dwarfs and TRAPPIST-1 because long-lived tidal heating is difficult to sustain against circularization in such compact systems \citep{hamer_ultra-short-period_2020,lee_carving_2025}, though induction heating could play a role \citep{kislyakova_electromagnetic_2018}. {\bf Bottom:} Heat redistribution for a fiducial 2~$M_\oplus$ super-Earth with nominal volatile inventories orbiting different stellar hosts, for zero and non-zero eccentricity, following \citet{koll_scaling_2022}, with observed planets from \citet{lin_persistent_2026}, including HD 3167 b from \citet{coy_evidence_2026} and LHS 1140 b from \citet{cherubim_helium_2026}. The airless valley boundaries encompass the possible thick volatile and/or silicate vapor atmospheres centered around the peridotite melting point ($T_\mathrm{eq}\approx1400\pm700$ K). Reflective silicate clouds ($A_B=0.70$) may further lower the thermal ratio. Machine-readable data reproducing the bottom-panel thermal ratio curves are provided as Data behind the Figure in the online article.}
\label{fig:cosmic_shoreline}
\end{figure*}

Figure \ref{fig:cosmic_shoreline} overlays the shorelines and sandbars from Figure \ref{fig:3x3_cosmic_shoreline} with additional simulations spanning a broader range of volatile inventories, eccentricities, and stellar types in escape velocity--cumulative XUV fluence space (top panels). The bottom panel shows calculations for heat redistribution for a fiducial 2~$M_\oplus$ super-Earth across four stellar types (G star, K star, M star, and TRAPPIST-1) at 5 Gyr using results from the simulations shown in Fig. \ref{fig:3x3_cosmic_shoreline} to constrain total atmospheric pressure. The heat redistribution model is used to calculate day-side temperature as a function of irradiation ($T_\mathrm{irr}$) and equilibrium ($T_\mathrm{eq}$) temperature, finding broad agreement with the observed population of rocky atmospheres, adopting values reported in \citet{lin_persistent_2026}. We include a case of mixed silicate vapor and thick volatile atmospheres with increased albedo from a reflective silicate cloud layer ($A_B=0.70$), demonstrating that a reflective silicate cloud layer can further reduce the thermal ratio on top of a thick volatile atmosphere. For a representative 2~$M_\oplus$ super-Earth orbiting M to G stars, our results suggest that planets with $T_\mathrm{eq} \gtrsim 600$--$700$~K are likely to reside in the airless valley, while those with $T_\mathrm{eq} \gtrsim 1900$~K, aided by tidal heating, or $T_\mathrm{eq} \gtrsim 2100$~K from stellar insolation alone, may be hot enough to enter the cosmic sandbar, where efficient interior-atmosphere exchange sustains a possible magma ocean-fueled volatile atmosphere. Our models predict, and are consistent with the data showing, a gradual increase in the dayside thermal ratio for hotter lava and magma ocean worlds with thick volatile atmospheres, because heat redistribution becomes less efficient at higher irradiation temperatures following the scaling of \citet{koll_scaling_2022}. We present polynomial best-fit scaling relations, same as Table \ref{tab:polyfit_au_mass}, but in escape velocity--cumulative XUV fluence space, for the time-evolving cosmic sandbars and shorelines as a function of planetary parameters in Table \ref{tab:polyfit_IXUV_ves}.

\section{Discussion}\label{sec:discussion}

We adopt the model of \citet{nguyen_effect_2026} as a baseline to explore the time-variable evolution of outgassing-escape dynamics governing the cosmic sandbar, shoreline, and airless valley for close-in rocky planets. However, significant uncertainties remain in (i.) the efficiency of thermal and non-thermal escape, (ii.) atmospheric collapse and heat redistribution at the cool edge of the shoreline, (iii.) poorly constrained initial volatile inventories and the possible role of H-rich primary atmospheres, (iv.) the long-term forcing from secular tidal excitation, (v.) vertical mixing and convection needed to resupply the upper atmosphere and regulate species-specific escape, (vi.) stellar XUV evolution histories that set the integrated escape budget, and (vii.) volcanic and tectonic replenishment, each of which we discuss in turn below, and results should be interpreted as a first-pass generalized framework for where rocky planets may harbor volatile atmospheres.


(i.) We focus on XUV-driven hydrodynamic escape as the dominant mass-loss mechanism \citep{owen_planetary_2012, owen_atmospheric_2019, catling_atmospheric_2017, pass_receding_2025, ji_cosmic_2025}. Jeans escape becomes relevant as XUV forcing reduces, with the transition from hydrodynamic outflow to Jeans-like escape occurring as the exobase Jeans parameter increases from $\lambda_{\rm esc}\sim2$--3 to $\lambda_{\rm esc}\gtrsim6$ \citep{volkov_thermally_2011}. The edge cases of cooler, more temperate close-in planets such as those in the TRAPPIST-1 system still receive $68$--$1982$ times Earth's EUV flux \citep{van_looveren_airy_2024}, consistent with hydrodynamic escape of light species when photolysis-derived H is supplied to the upper atmosphere, though heavier or strongly cooling thermospheres may remain closer to the Jeans-like regime \citep{tian_thermal_2009}. We therefore treat Jeans escape as a secondary process that broadens the shoreline transition but does not explicitly set the initial XUV-defined boundary for our model, nor does it act as the dominant escape mechanism for the sandbar. Additionally, we do not parameterize non-thermal escape processes such as ion pickup, sputtering, and photochemical escape, nor impact-driven atmospheric erosion, which may further thin or strip atmospheres interior to the shoreline and contribute to the observed population of bare rock worlds \citep{catling_atmospheric_2017, kegerreis_atmospheric_2020}. However, we expect them to play only a minor role in the sandbar regime, where XUV-driven hydrodynamic escape likely remains the dominant loss process.


(ii.) Quantifying atmospheric heat redistribution across the full range of close-in rocky planets, from hot lava worlds to cold collapsing atmospheres, is essential for interpreting thermal emission observations and distinguishing volatile atmospheres from bare rocks. For cold planets, inefficient day-night heat transport can drive nightside temperatures below condensation, leading to atmospheric collapse \citep{wordsworth_atmospheric_2015, auclair-desrotour_atmospheric_2020}. We adopt the analytic heat redistribution framework of \citet{koll_scaling_2022}, which relates the day-night temperature contrast to equilibrium temperature, surface pressure, and broadband longwave optical depth, providing a baseline translation from predicted atmospheric composition and pressure to redistribution efficiency. However, condensation and cold-trapping may reshape circulation and shift the cold-edge stability boundary of the cosmic shoreline \citep{ding_stabilization_2020, wordsworth_atmospheric_2015}, while at temperatures typical of lava worlds, short radiative timescales make redistribution predictions sensitive to assumed longwave optical depth, with magma-ocean heat transport unlikely to compensate for stellar forcing \citep{koll_temperature_2016, koll_scaling_2022, lai_ocean_2024, yang_ocean_2025}. Although the scalings of Koll (2022) were derived in the thin, dry limit, they are GCM-fitted to surface pressures of $\sim$100 bar and discussed there for hot, tidally locked lava worlds. We therefore adopt them as a first-order heat redistribution parameterization, noting that composition-dependent shortwave absorption, non-gray radiative transfer, and cloud opacity could modify $\varepsilon$ for thick volatile atmospheres \citep{shields_effect_2013,yang_stabilizing_2013,wordsworth_atmospheric_2015, hammond_linking_2017}.


(iii.) The volatile inventory available to form a secondary atmosphere depends not only on volatile inventory during bulk accretion but also on the largely unconstrained role of primordial hydrogen-rich envelopes, which may alter mantle volatile abundances through ingassing \citep{chachan_role_2018, modirrousta-galian_efficacy_2025} and redox reactions \citep{lichtenberg_redox_2021} during the magma ocean stage \citep{kite_water_2021, krissansen-totton_erosion_2024}. Additionally, initial volatile abundances are also expected to be shaped by a combination of early hydrodynamic escape processes, including catastrophic bolometric-heating-driven mass loss (known as `boil-off' \citep{owen_atmospheres_2016,tang_assessing_2024}) and subsequent XUV-driven escape \citep{lopez_how_2012, owen_kepler_2013, chen_evolutionary_2016, jin_compositional_2018, kubyshkina_overcoming_2018, rogers_photoevaporation_2021, tang_reassessing_2025} during the sub-Neptune evolution phase. As the primordial envelope depletes and its insulating effect diminishes, mantle pressure and temperature decline, driving outgassing of dissolved volatiles \citep{tang_reassessing_2025}. Because solubilities differ across species \citep{gupta_miscibility_2025}, the atmospheric composition evolves over time \citep{tang_hydrogen_2026}, which in turn regulates escape through changes in mean molecular weight, atmospheric structure, and thermal contraction, leading to preferential loss of species such as helium \citep{tang_hydrogen_2026}. A self-consistent treatment of the coupled sub-Neptune and super-Earth phases is required to determine initial volatile abundances, which we leave to future work.

We also do not include the role of core formation, even though metal-silicate differentiation can substantially modify the volatile budget available for later mantle outgassing by sequestering species such as H, C, and N into core-forming metal and by shifting mantle redox during magma ocean evolution \citep{hirschmann_magma_2012, grewal_speciation_2020, suer_distribution_2023, johansen_anatomy_2023}. As a result, our adopted initial mantle volatile inventory should be interpreted as an effective post-differentiation silicate reservoir, rather than a fully self-consistent outcome of coupled accretion, core partitioning, and atmospheric evolution.

Formation processes add further uncertainty to the initial volatile inventories of the rocky planets considered here. Around Sun-like stars, volatile budgets are shaped by stochastic planetesimal delivery, impact erosion, and subsequent core-mantle partitioning during late-stage accretion \citep{rubie_accretion_2015, sakuraba_impact_2019, sakuraba_numerous_2021, lock_atmospheric_2024}. In compact M dwarf systems, they may instead reflect a mixture of local accretion, pebble accretion, and inward migration from near or beyond the snowline \citep{ormel_formation_2017, coleman_pebbles_2019, schoonenberg_pebble-driven_2019, liu_pebble-driven_2020}. These pathways imprint large scatter, with models coupling accretion histories to volatile tracking, impact erosion, atmospheric loss, and pre-main-sequence evolution finding water inventories that span orders of magnitude across realizations and assumptions. Compact M dwarf planets may emerge either volatile-poor, if early irradiation, impact erosion, or inefficient delivery dominates, or volatile-rich, if they accrete near the snowline and migrate inward during pebble-driven growth \citep{tian_water_2015, alibert_formation_2017, ormel_formation_2017, ida_water_2019, schoonenberg_pebble-driven_2019, miguel_diverse_2020, kimura_predicted_2022, chen_impact_2022, childs_composition_2023, muller_formation_2024, chen_born_2025}. Fragile resonant chains, such as TRAPPIST-1's, further disfavor substantial late impact delivery, pointing instead to volatile accretion during pebble-driven growth \citep{raymond_upper_2022}. Our grids of ($F_{\mathrm{H_2O}}$, $F_{\mathrm{CO_2}}$) in Tables~\ref{tab:polyfit_au_mass} and~\ref{tab:polyfit_IXUV_ves} are therefore intended to span this formation-era scatter as effective post-accretion inventories, with JWST DDT detections and non-detections testing formation-set volatile inventories plus subsequent escape and outgassing together.


(iv.) We adopt the solid-viscoelastic tidal heating formulation of \citet{nguyen_effect_2026} and \citet{driscoll_tidal_2015}, assuming constant eccentricity as a simplification of Gyr-scale secular excitation that depends sensitively on system multiplicity, ordering, and packing \citep{bolmont_tidal_2013}. While liquid-layer tides can modify dissipation during solidification and the formulation does not explicitly model dynamical tides in a liquid magma ocean \citep{farhat_tides_2025, korenaga_tidal_2025}, a solid-body framework provides a reasonable approximation to the surface heat flux from tidal heating \citep{nguyen_effect_2026}. For temperate planets retaining surface oceans, ocean tidal dissipation may provide an additional heating channel \citep{shi_ocean_2025}. Asynchronous tides from spin-orbit coupling could contribute additional heating, though in general, synchronization and eccentricity circularization timescales are typically fast for short-period rocky planets ($\lesssim0.1$--3 Myr) unless secularly perturbed by companions in high-multiplicity systems \citep{rodriguez_spinorbit_2012, hansen_potentially_2015, brasser_long-term_2022, ferraz-mello_tidal_2025}.


(v.) Our XUV-driven monatomic escape model adopts an exobase-level, energy-limited hydrodynamic parameterization for atomic H, O, and C produced by photodissociation, implicitly assuming efficient resupply from below \citep{hunten_escape_1973, watson_dynamics_1981, owen_planetary_2012, catling_atmospheric_2017}. In reality, vertical transport across the homopause can bottleneck heavy species through diffusion and finite eddy mixing, making heavy-species loss supply-limited rather than purely energy-limited \citep{hunten_escape_1973, hunten_mass_1987, catling_atmospheric_2017, gronoff_atmospheric_2020, wordsworth_atmospheres_2022, modirrousta-galian_diffusion_2024}. We approximate these vertical-supply limitations by halting $\mathrm{CO_2}$-derived escape once $P_{\mathrm{H_2O}} \leq P_{\mathrm{CO_2}}$ or when the hydrogen flux falls below the crossover-mass drag threshold, while allowing $\mathrm{O_2}$ loss to continue whenever sufficient $\mathrm{H_2O}$ remains to sustain an H wind. This reflects that O is produced directly in the upper atmosphere via $\mathrm{H_2O}$ photolysis and is naturally co-located with the escaping H wind \citep{wordsworth_abiotic_2014, luger_extreme_2015}, whereas $\mathrm{CO_2}$-derived loss additionally requires upward transport from the surface and dissociation before removal. A complete treatment would resolve photolysis in the middle atmosphere, rather than at the exobase, and the photochemical partitioning among H, O, and C species and their recombination products during transport to the exobase \citep{hu_photochemistry_2012, tian_history_2015, wogan_jwst_2024}, including the inherently 3D chemistry-climate behavior of synchronously rotating planets \citep{chen_habitability_2019, chen_sporadic_2023, cooke_degenerate_2023, cooke_variability_2023}. Our gated fluxes should accordingly be read as an efficient-supply upper bound on escape. Future work should examine whether atomic O can recombine into molecular $\mathrm{O_2}$ and settle into the lower atmosphere, where it would be shielded from XUV and decoupled from the H wind, potentially stabilizing an abiotic $\mathrm{O_2}$-dominated atmosphere on longer timescales \citep{harman_abiotic_2015}.


(vi.) Recent work suggests cumulative XUV fluences for mid-to-late M dwarfs may be $\sim2$--3 times higher than canonical estimates once pre-main-sequence and flare contributions are included \citep{pass_receding_2025}, pushing many rocky planets toward the airless valley. While we adopt fiducial XUV parameterizations ($f_0$, $t_{\mathrm{sat}}$, $\beta_{\mathrm{XUV}}$) to represent generalized XUV histories across M--G stellar hosts, XUV histories remain highly uncertain on a star-by-star basis. We therefore show observed rocky close-in planets as an uncertainty range (gray vertical bars connecting planet points) in Figure~\ref{fig:cosmic_shoreline}, which span the conservative cumulative fluence estimates of \citet{pass_receding_2025} and our more canonical XUV implementation used in our evolution models. Relatedly, our stellar prescriptions isolate the time evolution of the integrated XUV and bolometric luminosities and do not resolve the wavelength-dependent spectral energy distribution, which shapes photolysis rates, thermospheric heating and cooling, and atmospheric chemistry, particularly for M dwarf hosts \citep{shields_effect_2013, eager-nash_implications_2020}, a coupling we leave to future work. Continued population-level constraints on M dwarf XUV histories, including long-baseline UV monitoring, improved flare frequency distributions, and XUV reconstructions from UV and X-ray proxies, will be essential for reducing this uncertainty, with observational constraints ideally extending into the fully convective M dwarf regime and eventually the substellar brown dwarf regime, where XUV histories and the possibility of exotic outgassing-escape dynamics remain largely unexplored \citep{barnes_habitable_2013, france_muscles_2016, gizis_k2_2017, bardalez_gagliuffi_ultracool_2019}.

(vii.) Post-solidification volcanic and tectonic outgassing could sustain either a substantial volatile or a tenuous stochastically replenished atmosphere on bare rock planets, making close-in exo-Venus and exo-Io analogs in the airless valley plausible \citep{kane_venus_2019, hill_smaller_2026}. Our model assumes no outgassing in the post-magma ocean phase, but future versions of our model will include this capability, allowing us to explore volcanic outgassing for those planets which retain small mantle volatile budgets. Within the airless valley, a volcanic island of transiently outgassed worlds may exist, sustained by tidal heating from system multiplicity and orbital resonances \citep{nicholls_volatile-rich_2026}. Recent transit spectroscopy of L 98-59 b \citep{bello-arufe_evidence_2025} and L 98-59 d \citep{banerjee_atmospheric_2024} tentatively reveals sulfur species, hinting at an actively outgassed atmosphere and suggesting such a population may already be within JWST's reach.

We also assume equilibrium volatile partitioning during magma ocean crystallization, with unconstrained entrainment of trapped melt into the solid mantle, so the retained interior volatile budget is controlled only by solid-melt partitioning. If instead a non-negligible fraction of volatile-rich melt is entrained and sequestered during crystallization, as explored in magma ocean crystallization studies, then more volatiles could be locked into the mantle against early outgassing, which would further favor later atmosphere retention and replenishment, potentially via volcanism or tectonic cycling \citep{bower_retention_2022, krissansen-totton_erosion_2024, sim_volatile_2024, lichtenberg_coupled_2026}.

Finally, we track only H$_2$O, CO$_2$, and photolytically produced O$_2$, and do not model N$_2$. On Venus and Mars, N$_2$ is a minor component ($\lesssim$3\% by mass) co-located with the dominant CO$_2$, so its contribution likely falls within the uncertainty of our modeled CO$_2$ inventories, while Earth's N$_2$-dominated atmosphere may partly reflect biological cycling and may not be representative of abiotic rocky planets \citep{stueken_evolution_2016, lammer_role_2019, spros_life_2021}. Titan's N$_2$-dominated atmosphere, likely inherited from accreted NH$_3$ ices, shows N$_2$ dominance does not require life, though its cold, volatile-rich origin may differ substantially from that of rocky close-in planets \citep{atreya_evolution_1978, mandt_protosolar_2014, johnson_evolution_2016}. Nitrogen melt solubility, speciation, and escape are also strongly redox and pressure dependent \citep{boulliung_oxygen_2020, grewal_speciation_2020, perez_whats_2025}. We provide the H--N binary diffusion coefficient in Table~\ref{tab:LJ_diffusion_values}, though extending the escape framework to nitrogen would require self-consistently coupling N-bearing melt speciation to photochemistry and magma ocean partitioning, which we leave to future work.


\section{Conclusions \& Summary}\label{sec:conclusion}

Applying the hydrodynamic escape and interior-atmosphere model of \citet{nguyen_effect_2026} with updated stellar evolution and XUV prescriptions \citep{ribas_evolution_2005, jackson_coronal_2012, baraffe_new_2015, schaefer_predictions_2016, birky_improved_2021, pass_receding_2025}, we examine atmospheric evolution across a wide range of stellar types and planetary interior conditions on rocky close-in planets, building upon previous cosmic shoreline frameworks \citep{zahnle_cosmic_2017, ji_cosmic_2025, van_looveren_habitable_2025} with the following main findings and implications:

\vspace{0.5\baselineskip}
\noindent (1.) \textit{Fuzzy Cosmic Shorelines, a New Cosmic Sandbar Regime and an Airless Valley}. Stellar-dependent equilibrium temperatures and XUV flux give rise to an outgassing-regulated cosmic sandbar and an escape-regulated cosmic shoreline, separated by an airless valley roughly centered around the peridotite melting point ($T_\mathrm{eq}\approx1400\pm700$ K). The sandbar is produced by long-lived magma ocean outgassing, which retains most volatiles in dissolution and exposes only a small atmospheric window to escape \citep{nicholls_self-limited_2025, nguyen_effect_2026}. By contrast, planets in the airless valley cool and solidify, forcing volatiles out of the melt and into the atmosphere, where they are more readily eroded because the solid phase stores volatiles much less efficiently than the melt. Both boundaries are sensitive to unconstrained volatile inventories \citep{schaefer_chemistry_2009, ji_cosmic_2025, perez_whats_2025} and stellar XUV evolution \citep{tu_extreme_2015, pass_receding_2025, van_looveren_habitable_2025}, and are likely further blurred by additional processes not modeled here, including tectonics \citep{foley_carbon_2018}, mantle convection \citep{maurice_onset_2017, boukare_timing_2018}, volcanism \citep{kite_exoplanet_2020, liggins_growth_2022}, impact erosion \citep{kegerreis_atmospheric_2020, wyatt_susceptibility_2020, krissansen-totton_erosion_2024}, dynamo shielding \citep{gunell_why_2018}, and Jeans escape \citep{tian_thermal_2009, nakayama_survival_2022, evans_thermally_2025}, precluding a definitive sharp boundary between airless and atmosphere-bearing planets.

\vspace{0.5\baselineskip}
\noindent (2.) \textit{Sandbar and Shorelines Evolve Across Myr-Gyr Timescales}. Atmospheric outcomes depend on time-evolving stellar-planetary parameters, where early XUV saturation drives aggressive escape during peak magma ocean outgassing on Myr timescales, while post-saturation decay and solidification shutoff set the final state within a few Gyr. Planetary mass sets a broad mass-dependent hydrodynamic fractionation transition that shapes the outer shoreline. Lower mass planets are more readily stripped, passing through mixed steam/O$_2$ states before becoming bare rock, while higher mass planets retain volatile atmospheres that can evolve from O$_2$-rich to lower pressure $\mathrm{CO_2}$-dominated residual states as $\mathrm{H_2O}$ depletion limits continued carbon escape and heavier $\mathrm{CO_2}$ is supplied less efficiently to the escaping region. This transition produces the inward shoreline bend near $\sim0.5$--$0.8M_\oplus$, consistent with the prediction that sub-Earth planets are more vulnerable to atmospheric loss \citep{hill_smaller_2026}.

\vspace{0.5\baselineskip}
\noindent (3.) \textit{Unlikely Sandbar for Cool Stars}. For a given orbital distance and planetary mass, the sandbar is most pronounced around K and G stars, where higher stellar effective temperatures and stellar masses provide greater insolation and tidal forcing to sustain the magma ocean, which in turn drives outgassing-regulated volatile atmospheres on close-in planets. By contrast, thick volatile atmospheres are unlikely around close-in and USP rocky planets orbiting M stars unless sustained by continuous secular tidal excitation on Gyr timescales or induction heating, which prolongs magma ocean cooling and outgassing sufficiently to overcome XUV-driven escape. A cosmic shoreline predicting secondary atmosphere retention remains possible for the largest and most temperate rocky planets orbiting M stars.

\vspace{0.5\baselineskip}
\noindent (4.) \textit{An Airless Valley Bridges the $I \propto v_{\mathrm{esc}}^{4}$ and $I \propto v_{\mathrm{esc}}^{3}$ Shoreline Scaling}. The observed scaling tensions between solar system and exoplanet populations, including disagreements in shoreline steepness \citep{zahnle_cosmic_2017, berta-thompson_3d_2025, meni-gallardo_empirical_2025}, can be reconciled with energy-limited hydrodynamic escape predictions \citep{zahnle_cosmic_2017, ji_cosmic_2025} by recognizing two distinct regimes, where lava worlds favor outgassing-regulated atmospheres while cooler temperate planets are dominated by escape post-solidification.

\vspace{0.5\baselineskip}
The boundary between airless worlds and those capable of sustaining an atmosphere, a prerequisite for surface habitability, remains one of the central open questions in planetary characterization. The ongoing JWST DDT survey and the forthcoming Roman and HWO missions will dramatically expand the population-level understanding of rocky planet atmosphere formation and evolution, underscoring the need for predictive frameworks that span diverse stellar and planetary regimes. Future work should address uncertainties in stellar XUV evolution and escape geometry, given the dependence of atmospheric escape on XUV and on day-nightside escape partitioning \citep{kang_escaping_2021, pass_receding_2025}, as well as post-solidification volatile replenishment and buffering through tectonics and volcanism and their dependence on stellar, planetary, and system-level dynamics \citep{noack_volcanism_2017, foley_carbon_2018, kite_exoplanet_2020, barth_magma_2021}, the role of atmosphere-ocean formation and evolution in regulating the long-term climate \citep{foley_habitability_2019, ding_multiple_2021}, and the cool-temperature limits of the cosmic shoreline where condensation and cold-trap collapse suppress atmospheric retention \citep{wordsworth_atmospheric_2015}.


\appendix
\twocolumngrid

\section{Cosmic Sandbar and Shoreline Best-Fit Relations}\label{app:fits}

Tables~\ref{tab:polyfit_au_mass} and~\ref{tab:polyfit_IXUV_ves} provide the polynomial best-fit scaling relations for the time-evolving cosmic sandbars and shorelines of Figures~\ref{fig:3x3_cosmic_shoreline} and~\ref{fig:cosmic_shoreline}, in planetary mass-semi-major axis space and escape velocity-cumulative XUV fluence space, respectively, as a function of stellar type, initial volatile inventory, eccentricity, and system age, for direct use in target selection and population-level comparisons.

\begin{deluxetable*}{lccc c ccc}
\tabletypesize{\scriptsize} 
\tablecaption{Polynomial best-fit scaling relations for cosmic sandbars and shorelines given by $\log_{10}(a/AU) = C_0 + C_1\,\log_{10}(M_p/M_\oplus) + C_2\,\bigl(\log_{10}(M_p/M_\oplus)\bigr)^2 + C_3\,\bigl(\log_{10}(M_p/M_\oplus)\bigr)^3$, in planetary mass-semi-major axis space. \label{tab:polyfit_au_mass}}
\tablewidth{1\columnwidth}
\tablehead{
\colhead{Stellar Type} &
\multicolumn{3}{c}{\textbf{Parameters}} &
\colhead{} &
\multicolumn{3}{c}{\textbf{Polynomial Regression Best-Fit}} \\
\cline{2-4}\cline{6-8}
\colhead{} &
\colhead{$F_{\mathrm{H_2O}}$} &
\colhead{$F_{\mathrm{CO_2}}$} &
\colhead{$e$} &
\colhead{\vsepHead} &
\colhead{Age (Gyr)} &
\colhead{Sandbar ($C_0$, $C_1$, $C_2$, $C_3$)} &
\colhead{Shoreline ($C_0$, $C_1$, $C_2$, $C_3$)}
}
\startdata
G           & $10^{-3}$ & $10^{-4}$ & 0    & \vsepDoubleTopGap & 0.01 & (-1.997, 2.542, -3.231, 1.289) & (-0.847, -0.370, 0.207, 0) \\
            &                    &                    &      & \vsepDouble & 0.10 & (-1.927, 0.933, -2.035, 1.291) & (-0.858, -0.355, 0.195, 0) \\
            &                    &                    &      & \vsepDouble & 1 & (-1.927, 0.933, -2.034, 1.291) & (-0.858, -0.355, 0.195, 0) \\
            &                    &                    &      & \vsepDouble & 5 & (-2.031, 0.829, -1.610, 0.961) & (-0.822, -0.253, 0.110, 0) \\
            & $10^{-3}$ & $10^{-4}$ & 0.01 & \vsepDoubleTopGap & 0.01 & (-1.640, 0.403, -0.049, 0) & (-0.845, -0.370, 0.207, 0) \\
            &                    &                    &      & \vsepDouble & 0.10 & (-1.702, 0.243, -0.106, 0) & (-0.847, -0.367, 0.204, 0) \\
            &                    &                    &      & \vsepDouble & 1 & (-1.702, 0.243, -0.106, 0) & (-0.847, -0.367, 0.204, 0) \\
            &                    &                    &      & \vsepDouble & 5 & (-1.794, 0.219, -0.064, 0) & (-0.824, -0.247, 0.106, 0) \\
            & $10^{-3}$ & $10^{-5}$ & 0    & \vsepDoubleTopGap & 0.01 & (-2.150, 2.381, -1.685, 0) & (-0.660, 0.125, 0.290, -0.470) \\
            &                    &                    &      & \vsepDouble & 0.10 & -- & (-0.270, -0.289, 0.363, -0.436) \\
            &                    &                    &      & \vsepDouble & 1 & -- & (-0.016, -0.994, 0.459, 0) \\
            &                    &                    &      & \vsepDoubleBottomGap & 5 & -- & (0.088, -0.960, 0.415, 0) \\
\hline
K           & $10^{-3}$ & $10^{-4}$ & 0    & \vsepDoubleTopGap & 0.01 & (-2.351, 3.347, -5.280, 2.595) & (-1.028, -0.325, 0.185, 0) \\
            &                    &                    &      & \vsepDouble & 0.10 & (-2.247, 1.051, -2.263, 1.338) & (-1.036, -0.319, 0.183, 0) \\
            &                    &                    &      & \vsepDouble & 1 & (-2.247, 1.051, -2.263, 1.338) & (-1.036, -0.319, 0.183, 0) \\
            &                    &                    &      & \vsepDouble & 5 & (-2.363, 0.915, -1.879, 1.058) & (-0.966, -0.286, 0.173, 0) \\
            & $10^{-3}$ & $10^{-4}$ & 0.01 & \vsepDoubleTopGap & 0.01 & (-1.707, 0.256, -0.165, 0) & (-1.027, -0.356, 0.173, 0) \\
            &                    &                    &      & \vsepDouble & 0.10 & (-1.742, 0.167, -0.113, 0) & (-1.043, -0.397, 0.269, 0) \\
            &                    &                    &      & \vsepDouble & 1 & (-1.666, 0.309, -0.227, 0) & (-0.765, -0.524, 0.106, 0) \\
            &                    &                    &      & \vsepDouble & 5 & (-1.842, 0.281, -0.157, 0) & (-0.962, -0.282, 0.155, 0) \\
            & $10^{-3}$ & $10^{-5}$ & 0    & \vsepDoubleTopGap & 0.01 & -- & (-0.852, 0.066, 0.380, -0.307) \\
            &                    &                    &      & \vsepDouble & 0.10 & -- & (-0.581, -0.102, 0.234, -0.294) \\
            &                    &                    &      & \vsepDouble & 1 & -- & (-0.172, -1.215, 0.701, 0) \\
            &                    &                    &      & \vsepDoubleBottomGap & 5 & -- & (-0.020, -1.188, 0.626, 0) \\
\hline
M           & $10^{-3}$ & $10^{-4}$ & 0    & \vsepDoubleTopGap & 0.01 & -- & (-1.407, -0.361, 0.233, 0) \\
            &                    &                    &      & \vsepDouble & 0.10 & -- & (-1.411, -0.364, 0.246, 0) \\
            &                    &                    &      & \vsepDouble & 1 & -- & (-1.221, -0.416, 0.074, 0) \\
            &                    &                    &      & \vsepDouble & 5 & -- & (-1.359, -0.258, 0.137, 0) \\
            & $10^{-3}$ & $10^{-4}$ & 0.01 & \vsepDoubleTopGap & 0.01 & (-1.955, 0.297, -0.048, 0) & (-1.405, -0.365, 0.209, 0) \\
            &                    &                    &      & \vsepDouble & 0.10 & (-1.952, 0.249, -0.131, 0) & (-1.413, -0.369, 0.258, 0) \\
            &                    &                    &      & \vsepDouble & 1 & (-1.948, 0.244, -0.126, 0) & (-1.221, -0.419, 0.064, 0) \\
            &                    &                    &      & \vsepDouble & 5 & (-2.058, 0.252, -0.127, 0) & (-1.355, -0.258, 0.130, 0) \\
            & $10^{-3}$ & $10^{-5}$ & 0   & \vsepDoubleTopGap & 0.01 & -- & (-1.218, -0.059, 0.117, -0.421) \\
            &                    &                    &     & \vsepDouble & 0.10 & -- & (-1.163, 0.301, 0.891, -1.266) \\
            &                    &                    &     & \vsepDouble & 1 & -- & (-0.746, -0.459, 0.095, 0) \\
            &                    &                    &     & \vsepDoubleBottomGap & 5 & -- & (-0.690, -0.358, 0.004, 0) \\
            & $10^{-2}$ & $10^{-3}$ & 0    & \vsepDoubleTopGap & 0.01 & -- & (\textit{sat.}) \\
            &                    &                    &      & \vsepDouble & 0.10 & -- & (-1.836, -0.421, 0.095, 0) \\
            &                    &                    &      & \vsepDouble & 1 & -- & (-1.708, -0.350, 0.059, 0) \\
            &                    &                    &      & \vsepDoubleBottomGap & 5 & -- & (-1.771, -0.282, 0.161, 0) \\
\hline
TRAPPIST-1  & $10^{-3}$ & $10^{-4}$ & 0   & \vsepDoubleTopGap & 0.01 & -- & (-1.637, -0.424, 0.291, 0) \\
            &                    &                    &     & \vsepDouble & 0.10 & -- & (-1.663, -0.443, 0.367, 0) \\
            &                    &                    &     & \vsepDouble & 1 & -- & (-1.672, -0.475, 0.421, 0) \\
            &                    &                    &     & \vsepDouble & 5 & -- & (-1.567, -0.272, 0.087, 0) \\
            & $10^{-3}$ & $10^{-4}$ & 0.01 & \vsepDoubleTopGap & 0.01 & (-2.048, 0.231, -0.142, 0) & (-1.660, -0.471, 0.288, 0) \\
            &                    &                    &      & \vsepDouble & 0.10 & (-2.042, 0.264, -0.087, 0) & (-1.694, -0.513, 0.352, 0) \\
            &                    &                    &      & \vsepDouble & 1 & (-2.039, 0.268, -0.083, 0) & (-1.705, -0.540, 0.407, 0) \\
            &                    &                    &      & \vsepDouble & 5 & (-2.150, 0.219, -0.120, 0) & (-1.567, -0.274, 0.093, 0) \\
            & $10^{-3}$ & $10^{-5}$ & 0    & \vsepDoubleTopGap & 0.01 & -- & (-1.207, 0.209, 0.100, -0.455) \\
            &                    &                    &      & \vsepDouble & 0.10 & -- & (-1.067, 0.078, 0.050, -0.384) \\
            &                    &                    &      & \vsepDouble & 1 & -- & (-0.987, -0.238, 0.001, 0) \\
            &                    &                    &      & \vsepDoubleBottomGap & 5 & -- & (-0.876, -0.319, 0.030, 0) \\
            & $10^{-2}$ & $10^{-3}$ & 0    & \vsepDoubleTopGap & 0.01 & -- & (\textit{sat.}) \\
            &                    &                    &      & \vsepDouble & 0.10 & -- & (-1.921, -0.362, 0.181, 0) \\
            &                    &                    &      & \vsepDouble & 1 & -- & (-1.919, -0.354, 0.178, 0) \\
            &                    &                    &      & \vsepDoubleBottomGap & 5 & -- & (-1.929, -0.271, 0.149, 0) \\
\enddata
\end{deluxetable*}

\begin{deluxetable*}{lccc c ccc}
\tabletypesize{\scriptsize} 
\tablecaption{Same as Table \ref{tab:polyfit_au_mass} but in $v_{\text{esc}}$-$I_{\mathrm{XUV,\oplus}}$ parameter space. Polynomial best-fit scaling relations for cosmic sandbars and shorelines in the form of $\log_{10}\left(I_{\mathrm{XUV,\oplus}}\right) = B_0 + B_1\log_{10}(v_{\mathrm{esc}}) + B_2\log_{10}(v_{\mathrm{esc}})^2 + B_3\log_{10}(v_{\mathrm{esc}})^3$. Times when the parameter space is saturated (sat.) with atmospheres have no defined shoreline or sandbar. Times when no sandbar exists are marked with an ellipsis. \label{tab:polyfit_IXUV_ves}}
\tablewidth{1\columnwidth}
\tablehead{
\colhead{Stellar Type} &
\multicolumn{3}{c}{\textbf{Parameters}} &
\colhead{} &
\multicolumn{3}{c}{\textbf{Polynomial Regression Best-Fit}} \\
\cline{2-4}\cline{6-8}
\colhead{} &
\colhead{$F_{\mathrm{H_2O}}$} &
\colhead{$F_{\mathrm{CO_2}}$} &
\colhead{$e$} &
\colhead{\vsepHead} &
\colhead{Age (Gyr)} &
\colhead{Sandbar ($B_0$, $B_1$, $B_2$, $B_3$)} &
\colhead{Shoreline ($B_0$, $B_1$, $B_2$, $B_3$)}
}
\startdata
G           & $10^{-3}$ & $10^{-4}$ & 0    & \vsepDoubleTopGap & 0.01 & (133.082, -290.566, 215.293, -53.016) & (-3.849, 8.545, -3.108, 0) \\
            &                    &                    &      & \vsepDouble & 0.10 & (104.046, -244.364, 197.597, -53.098) & (-3.543, 8.085, -2.927, 0) \\
            &                    &                    &      & \vsepDouble & 1 & (104.030, -244.333, 197.582, -53.098) & (-3.543, 8.085, -2.927, 0) \\
            &                    &                    &      & \vsepDouble & 5 & (80.988, -185.635, 148.517, -39.525) & (-1.626, 4.850, -1.651, 0) \\
            & $10^{-3}$ & $10^{-4}$ & 0.01 & \vsepDoubleTopGap & 0.01 & (6.405, -3.751, 0.736, 0) & (-3.853, 8.545, -3.108, 0) \\
            &                    &                    &      & \vsepDouble & 0.10 & (6.550, -4.669, 1.591, 0) & (-3.783, 8.434, -3.062, 0) \\
            &                    &                    &      & \vsepDouble & 1 & (6.550, -4.669, 1.591, 0) & (-3.783, 8.434, -3.062, 0) \\
            &                    &                    &      & \vsepDouble & 5 & (5.903, -3.215, 0.961, 0) & (-1.521, 4.691, -1.591, 0) \\
            & $10^{-3}$ & $10^{-5}$ & 0    & \vsepDoubleTopGap & 0.01 & (45.800, -66.100, 25.296, 0) & (-25.043, 72.221, -65.169, 19.331) \\
            &                    &                    &      & \vsepDouble & 0.10 & -- & (-27.794, 72.175, -61.865, 17.932) \\
            &                    &                    &      & \vsepDouble & 1 & -- & (-13.257, 19.899, -6.891, 0) \\
            &                    &                    &      & \vsepDoubleBottomGap & 5 & -- & (-12.544, 18.327, -6.230, 0) \\
\hline
K           & $10^{-3}$ & $10^{-4}$ & 0    & \vsepDoubleTopGap & 0.01 & (233.490, -536.706, 415.041, -106.731) & (-3.566, 7.606, -2.777, 0) \\
            &                    &                    &      & \vsepDouble & 0.10 & (110.658, -258.567, 207.102, -55.031) & (-3.482, 7.510, -2.747, 0) \\
            &                    &                    &      & \vsepDouble & 1 & (110.658, -258.567, 207.102, -55.031) & (-3.482, 7.510, -2.747, 0) \\
            &                    &                    &      & \vsepDouble & 5 & (90.488, -207.738, 165.107, -43.515) & (-3.268, 7.014, -2.597, 0) \\
            & $10^{-3}$ & $10^{-4}$ & 0.01 & \vsepDoubleTopGap & 0.01 & (6.909, -6.598, 2.477, 0) & (-3.548, 7.398, -2.597, 0) \\
            &                    &                    &      & \vsepDouble & 0.10 & (5.609, -4.473, 1.696, 0) & (-5.336, 10.645, -4.038, 0) \\
            &                    &                    &      & \vsepDouble & 1 & (8.155, -8.840, 3.408, 0) & (-3.931, 6.209, -1.591, 0) \\
            &                    &                    &      & \vsepDouble & 5 & (7.191, -6.483, 2.357, 0) & (-2.955, 6.426, -2.327, 0) \\
            & $10^{-3}$ & $10^{-5}$ & 0    & \vsepDoubleTopGap & 0.01 & -- & (-19.452, 53.260, -45.429, 12.627) \\
            &                    &                    &      & \vsepDouble & 0.10 & -- & (-17.932, 47.820, -41.555, 12.092) \\
            &                    &                    &      & \vsepDouble & 1 & -- & (-18.911, 28.729, -10.524, 0) \\
            &                    &                    &      & \vsepDoubleBottomGap & 5 & -- & (-17.821, 26.220, -9.398, 0) \\
\hline
M           & $10^{-3}$ & $10^{-4}$ & 0    & \vsepDoubleTopGap & 0.01 & -- & (-5.466, 9.314, -3.498, 0) \\
            &                    &                    &      & \vsepDouble & 0.10 & -- & (-5.689, 9.740, -3.693, 0) \\
            &                    &                    &      & \vsepDouble & 1 & -- & (-4.025, 4.609, -1.111, 0) \\
            &                    &                    &      & \vsepDouble & 5 & -- & (-3.666, 5.727, -2.057, 0) \\
            & $10^{-3}$ & $10^{-4}$ & 0.01 & \vsepDoubleTopGap & 0.01 & (4.051, -3.139, 0.721, 0) & (-5.096, 8.581, -3.138, 0) \\
            &                    &                    &      & \vsepDouble & 0.10 & (5.139, -5.489, 1.967, 0) & (-5.912, 10.145, -3.873, 0) \\
            &                    &                    &      & \vsepDouble & 1 & (4.523, -5.304, 1.892, 0) & (-3.877, 4.311, -0.961, 0) \\
            &                    &                    &      & \vsepDouble & 5 & (5.021, -5.380, 1.907, 0) & (-3.558, 5.507, -1.952, 0) \\
            & $10^{-3}$ & $10^{-5}$ & 0   & \vsepDoubleTopGap & 0.01 & -- & (-22.162, 61.133, -56.231, 17.315) \\
            &                    &                    &     & \vsepDouble & 0.10 & -- & (-73.062, 198.191, -177.188, 52.070) \\
            &                    &                    &     & \vsepDouble & 1 & -- & (-5.569, 5.506, -1.426, 0) \\
            &                    &                    &     & \vsepDoubleBottomGap & 5 & -- & (-3.383, 2.088, -0.060, 0) \\
            & $10^{-2}$ & $10^{-3}$ & 0    & \vsepDoubleTopGap & 0.01 & -- & (\textit{sat.}) \\
            &                    &                    &      & \vsepDouble & 0.10 & -- & (-2.674, 5.298, -1.426, 0) \\
            &                    &                    &      & \vsepDouble & 1 & -- & (-2.424, 3.775, -0.886, 0) \\
            &                    &                    &      & \vsepDoubleBottomGap & 5 & -- & (-3.376, 6.614, -2.417, 0) \\
\hline
TRAPPIST-1  & $10^{-3}$ & $10^{-4}$ & 0   & \vsepDoubleTopGap & 0.01 & -- & (-7.244, 11.486, -4.369, 0) \\
            &                    &                    &     & \vsepDouble & 0.10 & -- & (-8.556, 13.983, -5.509, 0) \\
            &                    &                    &     & \vsepDouble & 1 & -- & (-10.109, 15.858, -6.320, 0) \\
            &                    &                    &     & \vsepDouble & 5 & -- & (-3.249, 4.230, -1.306, 0) \\
            & $10^{-3}$ & $10^{-4}$ & 0.01 & \vsepDoubleTopGap & 0.01 & (4.490, -5.737, 2.132, 0) & (-7.418, 11.649, -4.324, 0) \\
            &                    &                    &      & \vsepDouble & 0.10 & (3.760, -4.186, 1.306, 0) & (-8.648, 13.894, -5.284, 0) \\
            &                    &                    &      & \vsepDouble & 1 & (3.215, -4.082, 1.246, 0) & (-10.186, 15.774, -6.110, 0) \\
            &                    &                    &      & \vsepDouble & 5 & (4.156, -4.978, 1.801, 0) & (-3.359, 4.430, -1.396, 0) \\
            & $10^{-3}$ & $10^{-5}$ & 0    & \vsepDoubleTopGap & 0.01 & -- & (-22.895, 63.743, -60.375, 18.714) \\
            &                    &                    &      & \vsepDouble & 0.10 & -- & (-19.735, 53.253, -50.438, 15.794) \\
            &                    &                    &      & \vsepDouble & 1 & -- & (-3.184, 1.336, -0.015, 0) \\
            &                    &                    &      & \vsepDoubleBottomGap & 5 & -- & (-3.960, 2.693, -0.450, 0) \\
            & $10^{-2}$ & $10^{-3}$ & 0    & \vsepDoubleTopGap & 0.01 & -- & (\textit{sat.}) \\
            &                    &                    &      & \vsepDouble & 0.10 & -- & (-4.504, 7.682, -2.717, 0) \\
            &                    &                    &      & \vsepDouble & 1 & -- & (-4.908, 7.544, -2.672, 0) \\
            &                    &                    &      & \vsepDoubleBottomGap & 5 & -- & (-3.543, 6.176, -2.237, 0) \\
\enddata
\end{deluxetable*}

\section{Binary Diffusion Coefficients from Lennard-Jones Chapman-Enskog Theory}\label{app:diffusion}

\begin{deluxetable*}{lcccccc} 
\tablecaption{Binary Diffusion Coefficients for Monatomic Species \label{tab:LJ_diffusion_values}} 
\tablewidth{1\columnwidth}
\scriptsize
\setlength{\tabcolsep}{3pt}
\setlength{\baselineskip}{0.90\baselineskip}
\tablehead{
  \colhead{Atomic Species} &
  \colhead{$b_i$ in H [$\mathrm{m}^{-1}\,\mathrm{s}^{-1}$]} & \colhead{Reference} &
  \colhead{$\sigma_{\mathrm{LJ}, i}$ [\AA]} &
  \colhead{Reference} &
  \colhead{$\epsilon_{\mathrm{LJ}}/k_B$ [K]} &
  \colhead{Reference}
}
\startdata
$\mathrm{H}$ & -- & -- & 2.05 & 1 & 145.0 & 1 \\
$\mathrm{C}$  & $b_{\mathrm{H,C}} = 8.31 \times 10^{19} T_{\text{esc}}^{0.67}$ & \textit{this work} & 3.298 & 1 & 71.4 & 1 \\
$\mathrm{N}$  & $b_{\mathrm{H,N}} = 8.31 \times 10^{19} T_{\text{esc}}^{0.67}$ & \textit{this work} & 3.298 & 1 & 71.4 & 1 \\
$\mathrm{O}$  & $b_{\mathrm{H,O}} = 1.00 \times 10^{20} T_{\text{esc}}^{0.67}$ & \textit{this work} & 2.75 & 1 & 80.0 & 1 \\
$\mathrm{O}$  & $b_{\mathrm{H,O}} = 4.8 \times 10^{19} T_{\text{esc}}^{0.75}$ & 2 & 2.75 & 1 & 80.0 & 1 \\
$\mathrm{He}$  & $b_{\mathrm{H,He}} = 1.53 \times 10^{20} T_{\text{esc}}^{0.66}$ & \textit{this work} & 2.551 & 3 & 10.22 & 3 \\
$\mathrm{Ne}$  & $b_{\mathrm{H,Ne}} = 1.11 \times 10^{20} T_{\text{esc}}^{0.66}$ & \textit{this work} & 2.820 & 3 & 32.8 & 3 \\
$\mathrm{Ar}$  & $b_{\mathrm{H,Ar}} = 6.93 \times 10^{19} T_{\text{esc}}^{0.68}$ & \textit{this work} & 3.33 & 1 & 136.5 & 1 \\
$\mathrm{Kr}$  & $b_{\mathrm{H,Kr}} = 5.67 \times 10^{19} T_{\text{esc}}^{0.69}$ & \textit{this work} & 3.655 & 3 & 178.9 & 3 \\
$\mathrm{Xe}$  & $b_{\mathrm{H,Xe}} = 4.56 \times 10^{19} T_{\text{esc}}^{0.69}$ & \textit{this work} & 4.047 & 3 & 231.0 & 3 \\
\enddata
\tablecomments{1. CANTERA \citep{goodwin_cantera_2015}; 2. \citet{zahnle_mass_1986}; 3. \citet{poling_properties_2001}.}
\end{deluxetable*}

This appendix details the analytic Lennard-Jones Chapman-Enskog calculation of the binary diffusion coefficients adopted in Section~\ref{sec:methods:escape}. Adopting the \citet{chapman_mathematical_1990} parameterization, as implemented by \citet{langenberg_technical_2020} and \citet{kobeissi_measurements_2025}, the binary diffusion coefficient between a species and hydrogen under Lennard-Jones (12-6) interactions is given by the Hirschfelder-Bird-Spotz (HBS) equation,

\begin{equation}\label{eqn:diffusion_coeff}
D_{\text{H},i} = \frac{3}{16\pi P_{\text{esc}} \sigma^{2}_{\text{H},i} \Omega^{(1,1)\ast}} \,
\sqrt{\dfrac{2\pi (k_B T_{\text{esc}})^3}{\mu_{\text{H},i}}}
\end{equation}

\noindent where $P_{\text{esc}}$ and $T_{\text{esc}}$ are the pressure and temperature at the escaping layer, $k_B$ is the Boltzmann constant, $\mu_{\text{H},i}$ is the reduced mass ($m_{\text{H},i}$) from \citet{hunten_escape_1973} defined as $\mu_{\text{H},i} = (m_\text{H} m_i)/(m_\text{H} + m_i)$, and $\sigma_{\text{H},i}$ is the Lennard-Jones binary collision diameter for hydrogen and heavier species $i$ defined as $\sigma_{\mathrm{H},i}=\tfrac{1}{2}\,(\sigma_{\mathrm H}+\sigma_i)$ via the Lorentz mixing rule. Note that $P_{\text{esc}}$ cancels out below in Eq. \ref{eqn:binary_diffusion} because $D_{\text{H},i}\propto P_{\text{esc}}^{-1}$ and $n_\text{esc}\propto P_{\text{esc}}$, so $b_{\text{H},i}=n_\text{esc}D_{\text{H},i}$ is pressure-independent. $\Omega^{(1,1)\ast}$ is the reduced diffusion collision integral \citep{neufeld_empirical_1972},

\begin{equation}\label{eqn:omega_collision}
\begin{aligned}
\Omega^{(1,1)\ast} &= \frac{1.06036}{{T^\ast_r}^{0.15610}}
+ \frac{0.19300}{\exp(0.47635\,T_{r}^\ast)} \\ 
&
+ \frac{1.03587}{\exp(1.52996\,{T^\ast_r})}
+ \frac{1.76474}{\exp(3.89411\,{T^\ast_r})} \,
\end{aligned}
\end{equation}

\noindent where $T_r^\ast = T_{\mathrm{esc}}/\epsilon_{\text{H},i,\mathrm{K}}$ is the reduced temperature, $\epsilon_{\text{H},i,\mathrm{K}} = \sqrt{\epsilon_{\text{H},\mathrm{K}}\,\epsilon_{i,\mathrm{K}}}$ is the mixed Lennard-Jones well-depth parameter in Kelvin (Berthelot rule), and $\sigma_{\text{H},i} = (\sigma_\text{H}+\sigma_i)/2$ is the mixed collision diameter (Lorentz rule), where $\epsilon_{\text{H},\mathrm{K}}$ and $\sigma_\text{H}$ are the potential well depth and kinetic diameter of hydrogen, and $\epsilon_{i,\mathrm{K}}$ and $\sigma_i$ are those of the heavier escaping species $i$, with all values tabulated in Table \ref{tab:LJ_diffusion_values}. 

The binary diffusion parameter $b_{\text{H},i}$, derived from the diffusion coefficient $D_{\text{H},i}$ via Equation (16) of \citet{hunten_escape_1973}, is thereby given as:

\begin{align}
b_{\text{H},i}
&= n_\text{esc}D_{\text{H},i}
 = \frac{P_{\text{esc}}}{k_B T_{\mathrm{esc}}}\,D_{\text{H},i} \notag\\
&= \frac{3}{16\,\pi\,\sigma_{\text{H},i}^{2}\,\Omega^{(1,1)\ast}}
\sqrt{\frac{2\pi k_B T_{\mathrm{esc}}}{\mu_{\text{H},i}}}
\label{eqn:binary_diffusion}
\end{align}

\noindent where $n_\text{esc}$ is the total number density at the escape level ($\text{m}^{-3}$) from the ideal gas law. We compute generalized binary diffusion forms $b_{\text{H},i}(T_\text{esc})$ in Table \ref{tab:LJ_diffusion_values} by evaluating Eq. \ref{eqn:binary_diffusion} over a rocky planet exobase temperature range ($T_\text{esc}=400$ to $3000$ K) and fitting the resulting values with a power law, noting that exobase temperatures vary strongly with atmospheric composition and radiative cooling efficiency \citep{catling_atmospheric_2017}.

\clearpage

\bibliography{2026_Cosmic_Sandbar}
\end{document}